\documentclass[preprint,12pt]{elsarticle}

\usepackage{amsmath,amssymb}
\usepackage{booktabs}
\usepackage{tikz}
\usetikzlibrary{arrows.meta,positioning,calc,fit,shapes.misc}
\usepackage{url}
\usepackage[hidelinks]{hyperref}
\usepackage{float}
\usepackage{tabularx}
\usepackage{algorithm}
\usepackage[noend]{algpseudocode}
\usepackage{longtable}
\usepackage{graphicx}
\usepackage{caption}
\usepackage{pgfplots}
\pgfplotsset{compat=1.18}
\usepgfplotslibrary{groupplots}

\newcommand{\lamlow}{\underline{\lambda}}
\newcommand{\lamhigh}{\overline{\lambda}}
\newcommand{\clearprice}{\rho}


\usepackage{xcolor}
\usepackage[normalem]{ulem}

\definecolor{suggestpurple}{HTML}{8338B5}
\definecolor{addblue}{HTML}{1769AA}
\definecolor{removeorange}{HTML}{D97706}

\newif\ifshowrevisions
\showrevisionsfalse     

\ifshowrevisions

  \newcommand{\suggestion}[1]{%
    \textcolor{suggestpurple}{%
      \textbf{[Suggestion:} #1\textbf{]}}}

  \newcommand{\addition}[1]{%
    \textcolor{addblue}{#1}}

  \newcommand{\removal}[1]{%
    \textcolor{removeorange}{\sout{#1}}}

\else

  \newcommand{\suggestion}[1]{}
  \newcommand{\addition}[1]{#1}
  \newcommand{\removal}[1]{}

\fi

\journal{Applied Energy}
\begin{document}
\begin{frontmatter}
\title{Verifiable Settlement and Pricing for Local Energy Markets: Architecture and Deployment on Hyperledger Besu}
\author[telecom]{Michele Fabi}
\author[ensae]{Baptiste Leloup}
\address[telecom]{T\'el\'ecom Paris and CREST, Palaiseau, France}
\address[ensae]{ENSAE Paris, Palaiseau, France}
\begin{abstract}
We implement privacy-preserving local electricity exchanges using an automated market maker on the Hyperledger Besu network under QBFT consensus. The proposed protocol operator collects individual electricity positions off-chain and publishes aggregate supply and demand on the blockchain, where smart contracts compute market prices every fifteen minutes. Individual positions remain off-chain; Poseidon2 commitments and UltraHonk zero-knowledge proofs, generated from Noir circuits, verify aggregation, payments, and daily balance updates. The architecture retains community coordination while making financial execution independently auditable. Signed receipts and member-triggered recourse prevent unverified or partial settlement. In a ten-household experiment using simulated US residential profiles, daily settlement uses approximately 19 million gas and produces eight-member proofs in 2.4 seconds. Over four simulated days, local trading covers 34.4\% of demand and reduces grid-only costs by 30.5\% under assumed tariffs. Numerically computed $\epsilon$-Nash Equilibrium electricity profiles capture 83–88\% of the social planner surplus. 
\end{abstract}

\begin{keyword}
Local energy market \sep automated market maker \sep collective self-consumption \sep blockchain \sep zero-knowledge proof \sep privacy
\end{keyword}
\end{frontmatter}

\section{Introduction}\label{sec:intro}

The promise of local electricity trading is not new, but realizing its economic value has proven difficult. In the 2010s, numerous initiatives attempted to build peer-to-peer energy networks, yet many struggled to gain traction~\cite{andoni2019,wu2022}.\footnote{Early peer-to-peer (P2P) energy trading startups and pilot projects have struggled to achieve large-scale commercialization due to significant regulatory friction and operational obstacles between decentralized prosumers and established grid operators~\cite{wu2022}. However, the sector has since matured significantly. Recent industry analyses value the core global P2P energy trading platforms market at \$138 million in 2024, projecting it to grow at a 29.7\% CAGR to reach \$2.34 billion by 2035~\cite{meticulous2025}. Broader estimates value the global P2P energy trading platforms market at \$556.2 million in 2024, scaling to \$10.4 billion by 2034 at a 34\% CAGR. Regionally, North America is leading this adoption, accounting for 36.2\% of the global market share in 2024~\cite{marketus2025}.} Major adoption obstacles were coordination costs and limited profitability: without a systematic way to monetize surplus generation, households had weaker incentives to invest in solar panels and shared infrastructure. Nevertheless, the gap between retail and feed-in tariffs creates an opportunity for mutually beneficial exchanges. Under net billing, a household exporting surplus power usually receives a modest feed-in tariff, while a neighbor purchases electricity at a higher retail rate. This differential includes more than suppliers' margins: retail tariffs may also recover network infrastructure costs, taxes, and other system charges. Yet, in some circumstances (for example, when they reduce upstream costs), coordinated exchanges may also be beneficial for the overall energy infrastructure.

Realizing the gains from electricity trading requires a mechanism that coordinates local electricity exchanges and determines how the resulting surplus is distributed among prosumers. Fabi et al.~\cite{fabi2025} address this market-design problem by introducing an automated market maker (AMM). In each fifteen-minute session, the mechanism aggregates the community's supply and demand and determines the prices received by producers and paid by consumers according to predefined pricing rules. These rules are derived from a set of economic principles (such as individual rationality and budget balance) and allocate the gains from the local exchanges within the bounds established by the grid tariffs. 

These economic properties define how the trading gains are allocated, but implementing the mechanism requires addressing practical considerations such as specifying arrangements for collecting electricity positions, calculating payments, updating accounts, and transferring funds. The question we address in this paper is how these operations can be carried out and independently checked without requiring households to rely solely on the organizing entity's financial calculations.

Our approach retains a community operator while changing how responsibilities are allocated within the community.\footnote{Depending on the jurisdiction, this role may be performed by a \emph{Personne Morale Organisatrice} (PMO) in France, a community solar program administrator in some US states, or a community energy manager or aggregator elsewhere. The legal duties of the community operator may vary across jurisdictions.} We adopt a hybrid architecture: The organizing entity operates under the applicable institutional arrangements and retains coordination, data-management, and compliance functions; households remain responsible for their own electricity decisions; and a shared ledger applies the agreed pricing rules and verifies the proposed financial settlement. The operator continues to collect individual positions and prepare the calculations in private, but the pricing formula used for exchanges is fixed in advance and any proposed settlement must pass the prescribed rules set by smart contracts. 

The first challenge our proposed approach faces concerns the confidentiality of household electricity data. A conventional on-chain implementation would require publishing each participant's electricity position at fifteen-minute intervals, exposing detailed consumption and generation patterns from which occupancy and daily routines could be inferred~\cite{chen2014}. The structure of the AMM, however, allows us to limit this disclosure. In each trading session, the volume matched within the community is the minimum of aggregate supply and demand, while the prices paid by buyers and received by sellers depend only on these aggregates and the applicable grid tariffs. Both quantities can therefore be computed on-chain without disclosing individual electricity positions. Although publishing aggregate data does not eliminate privacy risks, particularly in small communities, it avoids the direct disclosure of household-level electricity profiles.

Keeping individual electricity positions off-chain introduces a second challenge: ensuring that market outcomes remain independently verifiable. While the protocol smart contract infrastructure can compute prices directly from the published aggregates, it cannot determine from these aggregates alone whether they accurately reflect the positions recorded for each household. Nor can it verify the payments owed to or by individual participants, which depend on their private electricity positions and the prices determined by the mechanism. The protocol must therefore establish a verifiable link between the individual positions acknowledged by the operator, the aggregates used to determine market prices, and the resulting payments and account balances, without revealing the underlying household data.

We address these challenges through a an infrastructure implemented on Hyperledger Besu under QBFT consensus. The community operator collects individual electricity positions and returns digitally signed receipts, but publishes only aggregate supply and demand on-chain. Smart contracts compute the corresponding prices from predefined rules, while cryptographic commitments keep individual positions and account balances confidential. At the end of each day, zero-knowledge proofs generated from Noir circuits using UltraHonk verify that the committed positions reproduce the aggregates used to determine prices, and that individual payments and account balances are updated consistently with those prices.

The architecture also introduces safeguards against operator misbehavior and failure. Households can independently verify their commitments and request missing information through an on-chain recourse procedure. The protocol aggregates payments daily, allowing a twelve-hour window for verification and potential disputes. Once the required checks succeed, anyone may execute the financial transfers; otherwise, the protocol cancels them without modifying existing account balances. If the operator remains inactive for an extended period, households can independently recover their funds.

Finally, we evaluate the proposed infrastructure by executing complete
daily trading and payment cycles, including the generation and
on-chain verification of zero-knowledge proofs, on a
four-validator Hyperledger Besu network. For a community of
ten households, processing one day requires approximately
19 million gas units, while generating a proof for eight
households takes 2.4 seconds. We then assess the economic
performance of the three pricing rules using simulated
household consumption and photovoltaic generation profiles
and historical wholesale electricity prices. Under the
assumed tariff structure, local trading matches 34.4\% of
community demand and reduces the aggregate grid-only net
bill by 30.5\%. When households optimize
consumption and storage in response to the incentives generated by the pricing rule, the computed $\epsilon$-Nash equilibria
capture 83--88\% of the additional surplus achievable
through socially optimal coordination.

\section{Related literature and contribution}\label{sec:lit}

The contribution of this paper is to translate the  theoretical principles of local electricity market design into an operational and verifiable settlement mechanism. 

Distributed ledger technologies (DLTs) have long been explored for local electricity trading. One of the earliest and most prominent applications was the Brooklyn Microgrid, which experimented with blockchain-based peer-to-peer trading through a double-auction mechanism \cite{mengelkamp2018}. The experience also illustrated the regulatory and operational challenges of implementing decentralized electricity markets. More generally, double auctions introduce considerable complexity into markets that must operate at frequent intervals. These difficulties motivate interest in simpler, rule-based mechanisms that reduce the computational and coordination requirements of market clearing. At the infrastructure level, blockchain-based implementations have been extensively investigated \cite{andoni2019,christidis2021}.

The literature has proposed several alternatives to auction-based trading, including centralized optimization and pricing mechanisms based on aggregate supply and demand \cite{sousa2019,soto2021,capper2022}. Crowley et al. \cite{crowley2025}, for instance, formulate a bilevel optimization problem to determine individual electricity prices subject to budget balance and individual rationality. Zhou et al. \cite{zhou2018} evaluate the mid-market-rate and supply-demand-ratio pricing rules, while AMM-based implementations have been examined using Hyperledger Fabric \cite{kim2021}, storage-backed liquidity pools \cite{bevin2023}, and network-aware batch auctions \cite{saber2026}. Building on this literature, Fabi et al. \cite{fabi2025}  encompasses several pricing functions already used in experimental studies into a common theoretical framework, and proposes a method to construct new ones.

A central challenge in implementing local electricity markets concerns the confidentiality of household consumption and generation data. High-frequency smart-meter readings can reveal household activities \cite{chen2014}, while aggregating observations across participants does not eliminate the risk of inference, particularly in small communities \cite{buescher2017}. Several cryptographic approaches have been proposed to reconcile confidentiality with market operation. Secure multiparty computation and homomorphic encryption allow calculations over private inputs \cite{liu2024}, with protocols such as PA-Bill also incorporating mechanisms for accountable billing and dispute resolution \cite{erdayandi2023}. Zero-knowledge proofs offer an alternative means of verifying computations without disclosing their underlying inputs, and have been applied to confidential financial transactions \cite{zkledger,zether} and electricity trading \cite{hou2022}. Our approach exploits the aggregate-based structure of the AMM to compute prices on-chain while keeping individual electricity positions off-chain. In our implementation, cryptographic commitments and zero-knowledge proofs establish the consistency of these positions with published aggregates, individual payments, and account balances, while signed receipts and recourse procedures provide participants with means to challenge an operator that fails to cooperate. This preserves the verifiability of financial settlement without requiring the public disclosure of individual electricity profiles.

\section{Recap of pricing rules}\label{sec:mechanism}

We briefly recall the local energy market mechanism presented in
Fabi et al.~\cite{fabi2025}, before describing its implementation.
The mechanism allocates the profits of electricity exchanged within a
community. Energy is accounted for in consecutive fifteen-minute metering
intervals (96 per 24-hour day). In interval $t$, each participating household
or prosumer $n$ has a signed net energy position $x_{n,t}$, measured in
kilowatt-hours (kWh): $x_{n,t}>0$ denotes net injection and $x_{n,t}<0$ net
withdrawal. We write $s_{n,t}=(x_{n,t})^+$ for its supply and
$d_{n,t}=(-x_{n,t})^+$ for its demand, where $(z)^+=\max\{z,0\}$.
Thus each participant is a net seller, a net buyer, or inactive in a given
interval. We suppress the time index below when no confusion arises.

The relevant outside options are the tariffs at which a participant would
otherwise transact with its electricity supplier or other grid-facing
counterparty: an export remuneration (feed-in rate) $\lamlow$ and a retail
import tariff $\lamhigh$, expressed per kWh. We consider
$0\leq\lamlow<\lamhigh$. The spread $\lamhigh-\lamlow$ creates scope for mutually beneficial
local matching, subject to the applicable rules governing network charges and
energy sharing. Let
\[
  s=\sum_n s_n,\qquad d=\sum_n d_n,\qquad
  q=\min\{s,d\}.
\]
The quantity $q$ is matched within the community for settlement purposes;
any surplus $(s-d)^+$ is remunerated at $\lamlow$ and any shortfall
$(d-s)^+$ is billed at $\lamhigh$. In contrast to a bid-based auction,
the mechanism assigns uniform seller and buyer prices, $r(y)$ and $c(y)$,
from the aggregate supply--demand ratio $y=s/d$ and the tariff bounds.
Participant $n$ thus receives (or pays) $s_n r(y)-d_n c(y)$.
Both prices apply uniformly to all units supplied or demanded on the
corresponding side of the market.\footnote{If either $s$ or $d$ is zero, there is no
local matching and the non-zero side is settled at its grid tariff; the
formulas below concern $s,d>0$.}  To express the pricing rules compactly, define
\[
  u=\left(1-\frac{1}{y}\right)^+
    =\frac{(s-d)^+}{s},\qquad
  v=(1-y)^+=\frac{(d-s)^+}{d}.
\]
These are, respectively, the fractions of total supply exported to the grid
and total demand supplied from it; at most one is positive.

In the rest of the paper, we will  consider three pricing rules from~\cite{fabi2025} (and other experimental studies that used them before).

The \emph{mid-market rate} (MMR) rule~\cite{zhou2018} values the energy
matched locally at the arithmetic midpoint $\clearprice$ of the tariff band.
When the community is unbalanced, the net position of the community is averaged proportionally
across sellers or buyers, producing
\begin{equation}\label{eq:prices}
  \clearprice=\frac{\lamlow+\lamhigh}{2},\qquad
  c(y)=\clearprice+(\lamhigh-\clearprice)v,\qquad
  r(y)=\clearprice-(\clearprice-\lamlow)u.
\end{equation}
Thus $r=c=\clearprice$ in a balanced interval. Under excess demand, the
seller price remains at the midpoint and the buyer price includes the cost
of imports; under excess supply, the opposite adjustment applies.

The \emph{supply--demand ratio} (SDR) rule~\cite{zhou2018} instead makes the
seller price respond directly to the relative scarcity of local supply. In a
net-deficit interval, $r(y)$ decreases as $y$ approaches one, and the buyer
price is determined by the requirement that internal payments cover the grid
leg. When supply is sufficient to meet demand, both sides settle at the
feed-in rate:
\begin{equation}\label{eq:sdr}
  r(y)=\frac{\lamhigh\lamlow}
  {(\lamhigh-\lamlow)\min\{y,1\}+\lamlow},\qquad
  c(y)=
  \begin{cases}
    y\,r(y)+(1-y)\lamhigh,&y\leq1,\\
    \lamlow,&y>1.
  \end{cases}
\end{equation}

Lastly, the \emph{hyperbolic} rule derives prices from a shifted constant-product
trading function, rather than specifying them directly. Let $E$ and $M$
denote the energy and monetary accounting reserves of a notional pool. In
each interval, the curve is re-anchored at $(E,M)=(s,0)$ and constrained by
the tariff bounds. Its invariant is
\[
  \bigl(M+\sqrt{K\lamlow}\bigr)
  \bigl(E+\sqrt{K/\lamhigh}\bigr)=K,\qquad
  K=(s\chi)^2,\qquad
  \chi=\frac{\sqrt{\lamhigh\lamlow}}
  {\sqrt{\lamhigh}-\sqrt{\lamlow}}.
\]
Matching $q$ units leaves $E=s-q=(s-d)^+$ and determines the monetary
proceeds $M=s\,m(y)$ of the matched energy. Adding the export revenue or
import cost for the unmatched quantity, and sharing the resulting amounts
proportionally on each side, gives
\begin{equation}\label{eq:hyperbolic}
  \begin{aligned}
    m(y)&=\chi\left[
      \frac{\chi}{u+\chi/\sqrt{\lamhigh}}
       -\sqrt{\lamlow}\right],\\
    r(y)&=m(y)+u\,\lamlow,\qquad
    c(y)=y\,m(y)+v\,\lamhigh.
  \end{aligned}
\end{equation}
At $s=d$, the two prices coincide at
$\sqrt{\lamlow\lamhigh}$, the \textit{geometric mean} of the grid tariffs. As supply
exceeds demand, the curve reduces the value of additional local supply,
while the excess is remunerated at the feed-in rate.\footnote{The ``energy pool'' is an
accounting device for deriving prices, not an actual pool of energy stored or a separately
funded liquidity reserve. The implementation and theoretical characterization of adding an inventory to this mechanism is still an open problem.} 

All three rules depend on aggregate quantities rather than the identities or
order of individual transactions, and satisfy
$\lamlow\leq r(y)\leq c(y)\leq\lamhigh$. They are also exactly budget
balanced once the net grid exchange is included:
\begin{equation}\label{eq:mechanism-budget}
  d\,c(y)-s\,r(y)
  =\lamhigh(d-s)^+-\lamlow(s-d)^+.
\end{equation}
Consequently, for fixed electricity positions, the aggregate gain relative
to every participant trading at the grid tariffs is
\begin{equation}\label{eq:mechanism-gain}
  \underbrace{s\,[r(y)-\lamlow]}_{\text{sellers' gain}}
  +\underbrace{d\,[\lamhigh-c(y)]}_{\text{buyers' gain}}
  =(\lamhigh-\lamlow)\min\{s,d\}.
\end{equation}
The pricing rule therefore changes the \emph{distribution}, not the total
amount, of this gain. MMR splits it equally between buyers and sellers.
SDR assigns the full gain to buyers when local supply covers demand, while
its allocation varies with scarcity under excess demand. At balance, the
hyperbolic rule favors buyers relative to an equal split because the
geometric mean lies below the arithmetic mean.

We implement these rules through \textit{smart contracts}, programs
that blockchain nodes (or validators) execute according to predefined
conditions. Their execution consumes \emph{gas}, a measure
of computational and storage resources used by a transaction.
The community selects a pricing rule at deployment, and
the corresponding contract computes prices from the
fifteen-minute aggregates. The protocol nets the resulting
monetary obligations and settles them daily, without altering
the mechanism's session-level economic properties.

\section{Architecture} \label{sec:arch}

The pricing rules of Section~\ref{sec:mechanism} determine how a community allocates the gains from locally matched electricity. Implementing these rules raises a different question: how can participants verify the quantities, prices, and payments underlying their transactions without making their individual electricity profiles public? We address this question through a hybrid infrastructure that combines household-level data collection, an identified community operator, and a shared ledger that verifies financial settlement. This section develops the architecture from its design requirements to the roles of its participants, the ledger and smart contracts, and the procedures for recording and settling transactions. We call a household or prosumer registered in the local market a \emph{member}; its \emph{prosumer client} is the software through which it interacts with the protocol. Members are distinct from the \emph{validator nodes} that maintain the ledger.

\subsection{Requirements}\label{sec:requirements}

The market mechanism relies on two distinct levels of information.
While prices depend only on aggregate supply and demand in each
fifteen-minute session, calculating individual payments requires
each household's electricity position. This creates a tension
between confidentiality and verifiability. Publishing individual
positions on-chain would allow direct verification but expose
detailed consumption and generation profiles. Publishing only
aggregates allows prices to be computed on-chain but prevents
independent verification of the underlying quantities and
individual payments.

The architecture must therefore preserve the economic properties
established in Section~\ref{sec:mechanism} (tariff-bounded prices,
budget balance, and aggregate-based pricing) while making
individual financial calculations verifiable without disclosing
private positions. In particular, members must be able to verify
that the operator correctly aggregates their positions and
applies the resulting prices to their accounts.

To achieve this, the architecture combines three complementary
mechanisms:

\begin{itemize}
    \item Digitally \textit{signed receipts} and \textit{cryptographic
commitments} bind the operator to the positions it acknowledges.
\item \textit{Zero-knowledge proofs} establish the consistency of these
positions with published aggregates, individual payments, and
account balances without revealing private information.
\item  An on-chain \textit{recourse procedure} allows members to request
missing information and prevents financial transfers when
prescribed verification conditions remain unsatisfied.
\end{itemize}

Notice that the verification challenge we address concerns the processing of reported electricity
positions, not the authenticity of smart-meter measurements.
Detecting manipulated or inaccurate meter readings remains a
separate research problem~\cite{rial2011,wagh2024}, beyond the
scope of the proposed infrastructure. We therefore take reported
positions as inputs and focus on verifying their subsequent
processing.

\subsection{Roles and trust boundary}\label{sec:roles}

\begin{figure}[!htb]
\centering
\begin{tikzpicture}[
  font=\scriptsize, x=1cm, y=1cm,
  comp/.style={draw, rounded corners=2pt, align=center, inner sep=3pt, fill=white, text width=2.45cm, minimum height=0.7cm},
  contract/.style={draw, rounded corners=2pt, align=center, inner sep=3pt, fill=gray!12, text width=2.9cm, minimum height=0.7cm},
  grp/.style={draw, dashed, rounded corners=4pt, inner sep=7pt},
  flow/.style={-{Stealth[length=1.8mm]}, thick},
  back/.style={-{Stealth[length=1.8mm]}, thick, dashed},
  lbl/.style={fill=white, inner sep=1.2pt, align=center, font=\tiny}
]
  \node[comp] (meter) at (-0.3,4.6) {meter};
  \node[comp] (hems) at (-0.3,3.3) {local optimizer\\(battery, flexible load)};
  \node[comp] (client) at (-0.3,1.8) {prosumer client\\receipts, checks,\\recourse daemon};
  \node[grp, fit=(meter)(hems)(client)] (prem) {};
  \node[font=\scriptsize\bfseries, anchor=south west] at (prem.north west) {Prosumer premises ($\times N$)};

  \node[comp] (sessd) at (4.3,4.6) {session daemon\\aggregates, midnight freeze};
  \node[comp] (prover) at (4.3,3.3) {prover process\\one proof per batch};
  \node[comp] (datasvc) at (4.3,1.8) {data service\\signed receipts,\\encrypted answers};
  \node[grp, fit=(sessd)(prover)(datasvc)] (opg) {};
  \node[font=\scriptsize\bfseries, anchor=south west] at (opg.north west) {Community operator};

  \node[contract] (market) at (9.6,4.6) {Market contract\\cycle, recourse, settlement};
  \node[contract] (pricing) at (9.6,3.55) {Pricing library\\rules (\ref{eq:prices}), (\ref{eq:hyperbolic}), (\ref{eq:sdr})};
  \node[contract] (tariff) at (9.6,2.5) {Grid tariff contract\\$(\lamlow,\lamhigh)$ per session};
  \node[contract] (token) at (9.6,1.45) {Settlement token\\ERC-20, no issuance};
  \node[contract] (verif) at (9.6,0.4) {Verifier contracts\\built from the circuits};
  \node[grp, fit=(market)(pricing)(tariff)(token)(verif)] (ledg) {};
  \node[font=\scriptsize\bfseries, anchor=south west, align=left] at (ledg.north west) {Permissioned ledger\\(Besu, QBFT)};

  \node[comp, text width=2.9cm] (reporters) at (9.6,7.1) {tariff reporters\\quorum vote on each day's prices};
  \draw[flow] (reporters.east) -- ++(0.4,0) |- (tariff.east);

  \draw[dotted, thick] (6.8,7.95) -- (6.8,-1.55);
  \node[font=\tiny\itshape, anchor=east] at (6.7,7.8) {off-chain: private data and computation};
  \node[font=\tiny\itshape, anchor=west] at (6.9,7.8) {on-chain: shared, verified execution};

  \draw[flow] (meter) -- (hems);
  \draw[flow] (hems) -- (client);

  \draw[flow] ($(client.east)+(0,0.18)$) -- node[lbl, above]{net positions} ($(datasvc.west)+(0,0.18)$);
  \draw[back] ($(datasvc.west)+(0,-0.18)$) -- node[lbl, below]{signed receipts} ($(client.east)+(0,-0.18)$);

  \draw[flow] (sessd.east) -- node[lbl, above, pos=0.45]{$(s,d)$ every session} (market.west);
  \draw[flow] (prover.east) -- node[lbl, above, sloped, pos=0.42]{commitments, proofs} ($(market.west)+(0,-0.2)$);
  \draw[flow] (datasvc.east) -- node[lbl, below, sloped, pos=0.45]{data answers} ($(market.west)+(0,-0.35)$);

  \coordinate (below) at ($(ledg.south)+(0,-0.45)$);
  \draw[flow] (client.south) -- (client.south |- below)
    -- node[lbl, below, pos=0.4]{verifies every session, day and proof;\\files data requests, reveals, cancellations;\\triggers settlement} (below)
    -- (ledg.south);
\end{tikzpicture}
\caption{Actors and information flows}
\label{fig:architecture}
\end{figure}
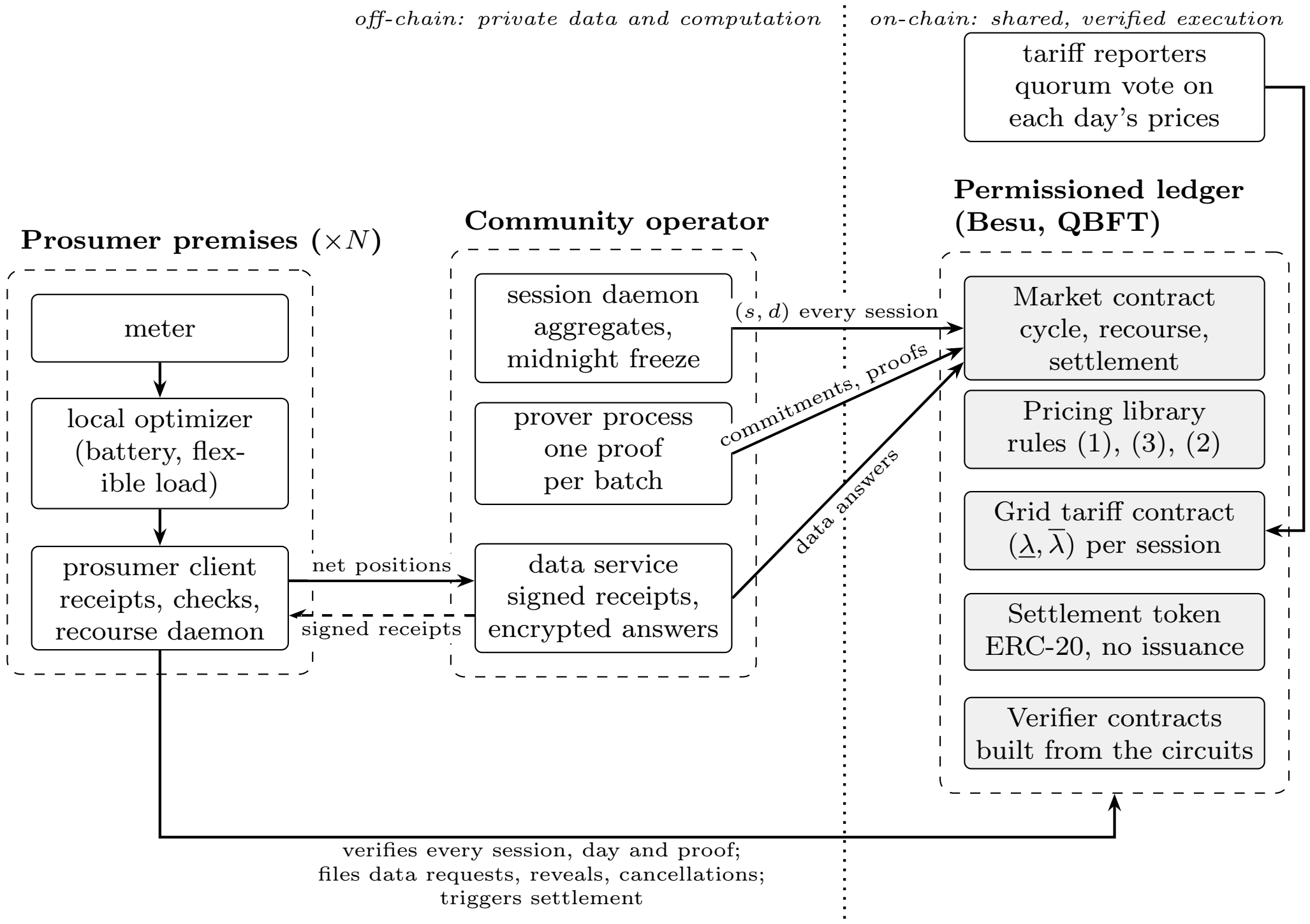

Figure~\ref{fig:architecture} organizes the elements of the proposed infrastructure into three groups. On the left, a household measures its electricity use and generation, chooses its net position, and communicates through a local prosumer client. The center shows the community operator, which receives individual positions and prepares the information required for market operation. On the right, the permissioned ledger records session aggregates, computes the corresponding prices, and verifies the proposed financial settlement. The arrows separate the private information exchanged between a household and the operator from the aggregate information and cryptographic evidence submitted to the ledger. We explain the responsibilities of each party below before turning to the underlying contracts.

\paragraph{The community operator}
The community operator connects members' individual electricity records to the shared market. Its role is necessary because the pricing mechanism relies on community-wide aggregates, while each household's underlying positions remain private. During each session, the operator acknowledges the positions it receives and submits their aggregate supply and demand to the ledger. At the end of the day, it prepares cryptographic evidence for the proposed settlement. Its proof-generating software, which we call the \emph{prover}, performs the private calculations needed to construct that evidence. 

The operator cannot choose a different pricing rule or authorize an unverified balance update: the market smart contract applies rules fixed at deployment and accepts settlement only after the prescribed checks. This arrangement constrains how the operator calculates and settles payments, without requiring participants to trust those calculations.

\paragraph{The prosumer client}
Each registered household uses a local software client to communicate with the operator and protect its interests in settlement. When the operator acknowledges a fifteen-minute electricity position, it returns a \emph{signed receipt}: a digitally signed record identifying the position and the session to which it belongs. The client retains these acknowledgments. 

At the daily close, the operator also sends a signed \emph{day-close packet} containing the random seed needed to reconstruct the commitment. The client combines that seed with its receipts and compares the result with the commitment published on the ledger. This comparison gives the household concrete, independently checkable evidence if the operator later reports a different position. 

The client also provides access to the data-request and settlement procedures discussed in Section~\ref{sec:cycle}; the household, rather than the software agent, remains the market member and holder of the associated rights.

\paragraph{The ledger}
The ledger provides a common record of market events that the operator cannot alter unilaterally. We implement its rules through \emph{smart contracts}: programs that validator nodes execute when authorized participants submit transactions. The \emph{market smart contract} records session aggregates and prices, coordinates the daily settlement, and enforces recourse conditions. It calls other contracts to obtain tariff bounds, verify cryptographic proofs, and transfer settlement tokens. 

Rather than publishing the numerical amounts held in each member's account, the ledger stores cryptographic commitments to these balances. We describe how these components interact in Sections~\ref{sec:ledger} and~\ref{sec:contracts}.

\paragraph{The tariff reporters}
The market also needs the grid tariffs that establish each session's external buying and selling prices. The tariff reporters supply these reference prices independently of the community operator. Under a predetermined tariff, a grid administrator publishes the applicable schedule; under a variable tariff, several authorized reporters submit daily tariff vectors, and the grid tariff smart contract accepts a vector when a specified quorum agrees. Separating the reporting of tariffs from the reporting of household positions prevents the community operator from controlling both inputs to the pricing calculation.

Table~\ref{tab:threat} summarizes the safeguards and remaining dependencies associated with each participant. It distinguishes errors or deviations that the market smart contract can prevent from conduct that requires a separate institutional or metering safeguard.

\begin{table}[H]
\centering
\caption{Trust boundaries}
\label{tab:threat}
\scriptsize
\setlength{\tabcolsep}{4pt}
\renewcommand{\arraystretch}{1.05}
\setlength{\aboverulesep}{1.5pt}
\setlength{\belowrulesep}{1.5pt}
\begin{tabularx}{\textwidth}{@{}>{\raggedright\arraybackslash}p{2.0cm}>{\raggedright\arraybackslash}X>{\raggedright\arraybackslash}X@{}}
\toprule
Actor & Protocol safeguards & Remaining dependencies \\
\midrule

Community operator and prover &
\textbullet\ Fixed pricing rule; proof verification before settlement.\par
\textbullet\ Member verification of receipts against daily commitments.\par
\textbullet\ On-chain record of unanswered requests. &
\textbullet\ Operator access to private positions.\par
\textbullet\ Potentially incorrect meter readings or reported positions.\\

\midrule[0.3pt]

Validator nodes &
\textbullet\ QBFT finality under fault assumptions.\par
\textbullet\ Member-initiated settlement actions. &
\textbullet\ Block production requires a quorum.\par
\textbullet\ Validator timing assumptions.\\

\midrule[0.3pt]

Tariff reporters &
\textbullet\ Quorum agreement on identical tariff vectors (feed mode). &
\textbullet\ Incorrect tariffs from inaccurate or colluding quorum.\\

\midrule[0.3pt]

Floor administrator &
\textbullet\ Operator proposal and administrator approval for floor changes. &
\textbullet\ Potential collusion to exclude members.\\

\midrule[0.3pt]

Households and permitted observers &
\textbullet\ Public aggregates and commitments, not individual positions.\par
\textbullet\ Cancellation restricted to specified grounds. &
\textbullet\ Inference risks from aggregates and external data.\\

\bottomrule
\end{tabularx}
\end{table}

\subsection{Hyperledger Besu under QBFT}
\label{sec:ledger}

Verifiable financial settlement requires a shared and final
record of the aggregates, prices, and commitments submitted
throughout the trading day. We implement this record on
Hyperledger Besu, which combines Ethereum-compatible smart
contracts with a configurable permissioned network.

\paragraph{Consensus and validators}
 We adopt Quorum Byzantine Fault Tolerance (QBFT) consensus~\cite{qbft},
under which identified \emph{validators} propose and agree on
blocks of transactions. The community determines the number
and identity of validators according to its governance and
fault-tolerance requirements.

\paragraph{Permissioned access}
The choice of a permissioned network also reflects the
institutional structure of local electricity markets. Collective
self-consumption arrangements already involve identified
participants within a defined geographical perimeter and
therefore do not require permissionless validation. Moreover,
a public blockchain would expose aggregate electricity profiles
and financial activity to unrestricted observation. A permissioned
network instead restricts participation in ledger maintenance
to authorized entities, potentially including the distribution
system operator and independent community representatives.
This reduces unnecessary data exposure without eliminating
inference risks for those with access to the ledger.

\paragraph{Finality and timekeeping}
Beyond governance and confidentiality, QBFT provides
\emph{finality}: once validators commit a block under the
protocol's consensus assumptions, its transactions cannot be
replaced by a competing history.\footnote{This is not generally the case in all blockchains, such as those based on Nakamoto consensus.} This property ensures that
daily financial calculations rely on previously agreed prices
and commitments. 

Block timestamps also provide the common
clock used by the market contract. Given a timestamp $\tau$,
the contract identifies the session as
$\lfloor\tau/900\rfloor\bmod96$ and the day as
$\lfloor\tau/86400\rfloor$. Unlike block-height-based approaches
\cite{christidis2021}, this convention follows elapsed time,
making validator synchronization relevant to the protocol's
operation.

\paragraph{Network configuration}
Finally, the network configuration must accommodate the
computational requirements of cryptographic verification and
daily transfers. The contract-size limit must support the
generated proof verifiers (approximately 23.7~kB each), while
the block gas limit must permit atomic daily settlement.
Transaction fees must also remain compatible with members'
information and recourse rights. We therefore configure the
test network with a zero minimum gas price, allowing members
to invoke these procedures without holding a separate native
cryptocurrency, and a five-second block interval, well below
the fifteen-minute trading frequency.

\subsection{Smart contracts}\label{sec:contracts}

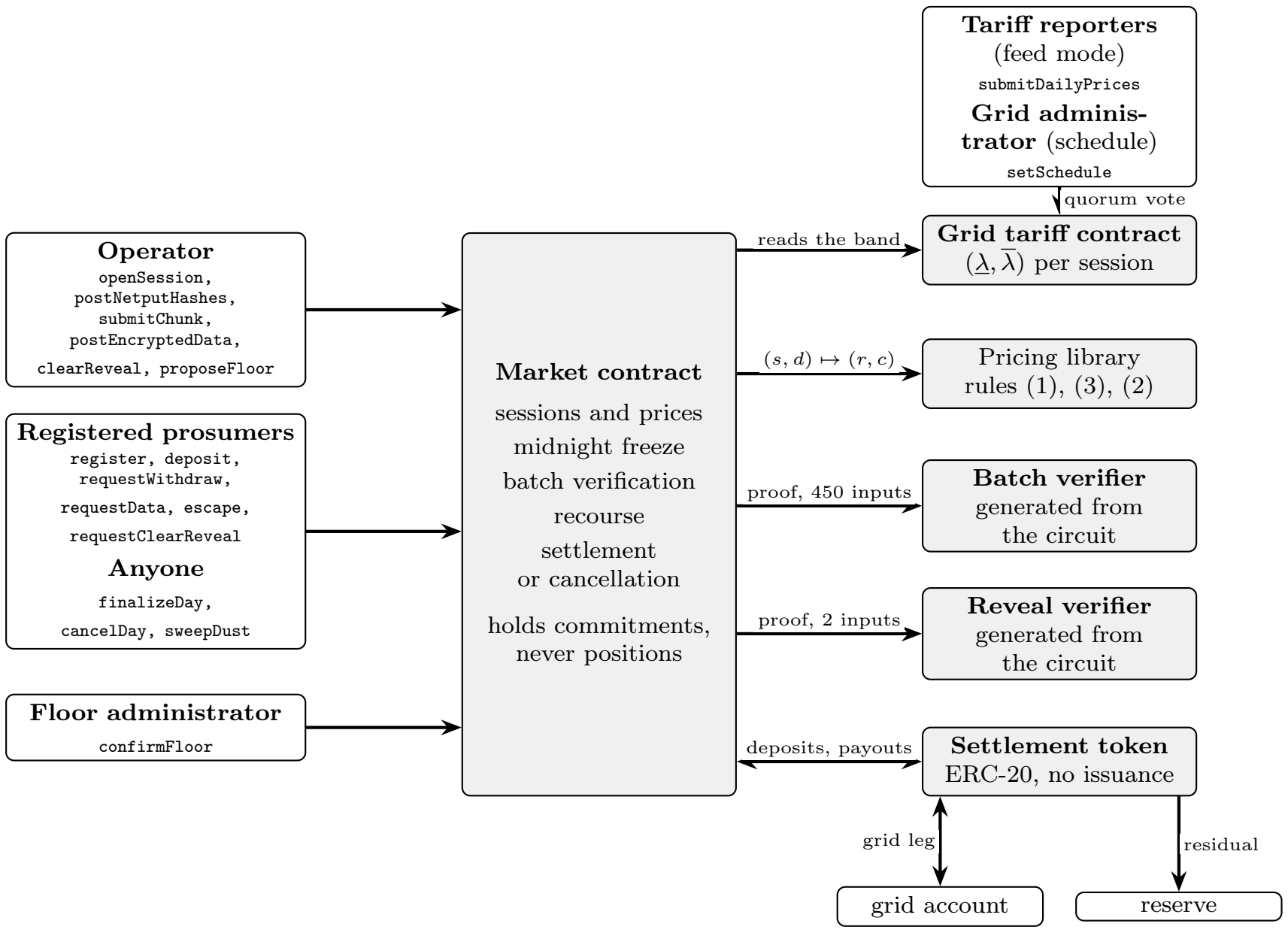
\begin{figure}[H]
\centering
\begin{tikzpicture}[
  scale=0.75, transform shape,
  font=\scriptsize, x=1cm, y=1cm,
  caller/.style={draw, rounded corners=2pt, align=center, inner sep=3pt, fill=white, text width=3.3cm},
  contract/.style={draw, rounded corners=2pt, align=center, inner sep=3pt, fill=gray!12, text width=3.0cm, minimum height=0.8cm},
  lib/.style={contract, dashed},
  acct/.style={draw, rounded corners=2pt, align=center, inner sep=3pt, fill=white, text width=2.2cm},
  flow/.style={-{Stealth[length=1.8mm]}, thick},
  both/.style={{Stealth[length=1.8mm]}-{Stealth[length=1.8mm]}, thick},
  lbl/.style={fill=white, inner sep=1.2pt, align=center, font=\tiny},
  fn/.style={font=\tiny\ttfamily}
]
  \node[caller] (op) at (0,4.3) {\textbf{Operator}\\[1pt]
    {\tiny\ttfamily openSession, postNetputHashes,\\ submitChunk, postEncryptedData,\\ clearReveal, proposeFloor}};
  \node[caller] (mem) at (0,1.7) {\textbf{Registered prosumers}\\[1pt]
    {\tiny\ttfamily register, deposit, requestWithdraw,\\ requestData, escape, requestClearReveal}\\[2pt]
    \textbf{Anyone}\\[1pt]
    {\tiny\ttfamily finalizeDay, cancelDay, sweepDust}};
  \node[caller] (fa) at (0,-0.6) {\textbf{Floor administrator}\\[1pt]
    {\tiny\ttfamily confirmFloor}};

  \node[contract, text width=3.0cm, minimum height=6.6cm] (market) at (5.2,1.9) {\textbf{Market contract}\\[4pt]
    sessions and prices\\[2pt] midnight freeze\\[2pt] batch verification\\[2pt] recourse\\[2pt] settlement\\ or cancellation\\[6pt]
    holds commitments,\\ never positions};

  \node[contract] (tariff) at (10.6,5.0) {\textbf{Grid tariff contract}\\ $(\lamlow,\lamhigh)$ per session};
  \node[contract] (pricing) at (10.6,3.55) {Pricing library\\rules (\ref{eq:prices}), (\ref{eq:hyperbolic}), (\ref{eq:sdr})};
  \node[contract] (dayv) at (10.6,2.0) {\textbf{Batch verifier}\\ generated from the circuit};
  \node[contract] (revv) at (10.6,0.5) {\textbf{Reveal verifier}\\ generated from the circuit};
  \node[contract] (token) at (10.6,-1.0) {\textbf{Settlement token}\\ ERC-20, no issuance};

  \node[caller, text width=3.0cm] (rep) at (10.6,6.8) {\textbf{Tariff reporters} (feed mode)\\ {\tiny\ttfamily submitDailyPrices}\\[1pt]
    \textbf{Grid administrator} (schedule)\\ {\tiny\ttfamily setSchedule}};
  \draw[flow] (rep.south) -- node[lbl, right]{quorum vote} (tariff.north);

  \node[acct] (gridacc) at (9.2,-2.7) {grid account};
  \node[acct] (reserve) at (12.0,-2.7) {reserve};
  \draw[both] (token.south -| gridacc.north) -- node[lbl, left]{grid leg} (gridacc.north);
  \draw[flow] (token.south -| reserve.north) -- node[lbl, right]{residual} (reserve.north);

  \draw[flow] (op.east) -- (market.west |- op.east);
  \draw[flow] (mem.east) -- (market.west |- mem.east);
  \draw[flow] (fa.east) -- (market.west |- fa.east);

  \draw[flow] (market.east |- tariff.west) -- node[lbl, above]{reads the band} (tariff.west);
  \draw[flow] (market.east |- pricing.west) -- node[lbl, above]{$(s,d)\mapsto(r,c)$} (pricing.west);
  \draw[flow] (market.east |- dayv.west) -- node[lbl, above]{proof, 450 inputs} (dayv.west);
  \draw[flow] (market.east |- revv.west) -- node[lbl, above]{proof, 2 inputs} (revv.west);
  \draw[both] (market.east |- token.west) -- node[lbl, above]{deposits, payouts} (token.west);
\end{tikzpicture}
\caption{Smart contract ecosystem.}
\label{fig:contracts}
\end{figure}

The ledger executes the market through a small set of interoperating smart contracts. As Figure~\ref{fig:contracts} shows, the \emph{market smart contract} is the entry point for transactions from the operator and registered households. It calls the grid tariff smart contract to obtain the reference prices, the pricing library to calculate the two market prices, and specialized verifier contracts to check cryptographic proofs. An ERC-20 settlement token (a standardized digital asset managed by a separate token contract) implements monetary transfers~\cite{erc20}. We describe each component below, beginning with the pricing calculation and then following the daily settlement workflow.

\paragraph{The pricing library}
The pricing library computes the tariff-bounded prices defined in Section~\ref{sec:mechanism}. Unlike an automated market maker that maintains a pool of tradable assets, this implementation holds no inventory: it applies a fixed pricing formula directly to the session's aggregate supply $s$, aggregate demand $d$, and external tariff bounds $(\lamlow,\lamhigh)$. The library neither transfers funds nor chooses prices independently of these inputs.\footnote{The implementation uses USD as its accounting denomination. The protocol can use other currencies with corresponding token and accounting parameters.} The market smart contract fixes the rule when the community deploys the system, and every validator can reproduce the resulting seller and buyer rates $(r,c)$.

To compute these prices on a blockchain, the library uses fixed-point integer arithmetic instead of unrestricted real-number calculations. It represents electricity in watt-hours as unsigned 32-bit integers and prices at a precision of $10^{-9}$~USD per kilowatt-hour; multiplying the two yields the internal monetary unit of $10^{-12}$~USD. The mid-market and supply--demand-ratio rules use elementary arithmetic, while the hyperbolic rule also requires approximating square roots. The library performs intermediate computations at 18-decimal precision, after which the market smart contract rounds seller prices down and buyer prices up. Section~\ref{sec:choices} explains why this conservative rounding preserves budget balance. The explicit formulas appear in Section~\ref{sec:mechanism}; we omit their direct pseudocode translation here.

\paragraph{The market contract}
The market contract coordinates the daily lifecycle. It first records the session aggregates and prices, then freezes the relevant account state at day close, verifies the proofs submitted for each batch, and finally either applies all eligible balance updates or cancels the proposed settlement. 

The market contract also enforces participants' information rights and provides an independent exit procedure after fourteen days without successful settlement. To keep member-level information private, it stores commitments rather than electricity positions or account balances as publicly readable numerical amounts. Balances use unsigned 64-bit integers in the internal unit (approximately eighteen million US dollars per member), and the circuits enforce the same range. Algorithms~\ref{alg:market-day} and~\ref{alg:market-settle} formalize the ordinary transition from session recording to finalization or cancellation. Section~\ref{sec:cycle} explains the recourse and recovery actions in the order members encounter them.
\begin{algorithm}[H]
\caption{Market contract, recording and proving day $e$}\label{alg:market-day}
\scriptsize\linespread{0.93}\selectfont
\begin{algorithmic}[1]
\Require Batch size $Q = 8$, proof window $W = 12$~h, pricing library and
tariff contract fixed at deployment, block timestamp $\tau$.
\Ensure The prices of every session of day $e$, the frozen state of the day,
and the verified batches with their staged results.
\Statex \textbf{(i) Session opening}, called by the operator during day $e$
\Procedure{openSession}{$e, t, s, d$}
  \State \textbf{require} session $t$ of day $e$ is the one in progress at $\tau$ and not yet opened
  \State $(\lamlow, \lamhigh) \gets$ \Call{Tariff.getPrices}{$\tau$}
  \State $(r, c) \gets$ \Call{Pricing}{$s, d, \lamlow, \lamhigh$} \Comment{pricing rules of Section~\ref{sec:mechanism}}
  \State record $s_t, d_t$ and $\lfloor r \rfloor, \lceil c \rceil, \lceil \lamlow \rceil, \lfloor \lamhigh \rfloor$ \Comment{rounded against the market}
\EndProcedure
\Statex \textbf{(ii) Midnight freeze}, called by the operator after day $e$
\Procedure{postNetputHashes}{$e, C_1, \dots, C_N$}
  \State \textbf{require} day $e$ is over, is the next one to close, and the previous day is not closing
  \For{$n = 1$ \textbf{to} $N$}
    \State store $C_n$, freeze the deposit, withdrawal request and floor commitment of $n$
  \EndFor
  \State $N_e \gets N$, \quad $J \gets \lceil N_e / Q \rceil$, \quad deadline $\gets$ end of day $e + W$
  \State state $\gets$ Closing
\EndProcedure
\Statex \textbf{(iii) Batch verification}, callable by anyone
\Procedure{submitChunk}{$e, j$, outputs, $\pi$}
  \State \textbf{require} state is Closing and batch $j < J$ is not yet verified
  \State $\xi \gets$ public inputs read from storage and from the claimed outputs \Comment{450 field elements}
  \State \textbf{require} \Call{Verifier.verify}{$\pi, \xi$}
  \For{$t = 1$ \textbf{to} $96$}
    \State $\Sigma^s_t \gets \Sigma^s_t + s^{(j)}_t$, \quad $\Sigma^d_t \gets \Sigma^d_t + d^{(j)}_t$
  \EndFor
  \State add $\mathrm{out}^{(j)}$ and $\mathrm{in}^{(j)}$ to the day's totals
  \State stage the new balance commitments and paid withdrawals of batch $j$
\EndProcedure
\end{algorithmic}
\end{algorithm}

\begin{algorithm}[H]
\caption{Market contract, settling or canceling day $e$}\label{alg:market-settle}
\scriptsize\linespread{0.93}\selectfont
\begin{algorithmic}[1]
\Require The state of day $e$ after Algorithm~\ref{alg:market-day}, grace
period $\Gamma = 6$~h, block timestamp $\tau$.
\Ensure Day $e$ is either settled once, as the recorded prices imply, or
canceled with nothing moved.
\Statex \textbf{(iv) Settlement}, callable by anyone
\Procedure{finalizeDay}{$e$}
  \State \textbf{require} all $J$ batches verified, $\tau \ge$ deadline, no data request open
  \For{$t = 1$ \textbf{to} $96$}
    \State \textbf{require} $\Sigma^s_t = s_t$ and $\Sigma^d_t = d_t$ \Comment{reconciliation (\ref{eq:reconcile})}
  \EndFor
  \State \textbf{require} $\sum_j \mathrm{out}^{(j)} = \sum_t r_t s_t$ and $\sum_j \mathrm{in}^{(j)} = \sum_t c_t d_t$
  \State $\text{grid}^- \gets \sum_t \lamhigh_t\,(d_t - s_t)^+$, \quad $\text{grid}^+ \gets \sum_t \lamlow_t\,(s_t - d_t)^+$
  \State residual $\gets \bigl(\sum_j \mathrm{in}^{(j)} + \text{grid}^+\bigr) - \bigl(\sum_j \mathrm{out}^{(j)} + \text{grid}^-\bigr)$ \Comment{$\ge 0$ by rounding}
  \State apply the staged commitments, consume the frozen queues, pay the withdrawals
  \State settle the grid leg, \quad state $\gets$ Finalized
\EndProcedure
\Statex \textbf{(v) Cancellation}, callable by anyone
\Procedure{cancelDay}{$e$}
  \State \textbf{require} proofs missing after the deadline, \textbf{or} a data request unanswered past its own deadline, \textbf{or} no settlement by deadline $+\,\Gamma$ with no request open
  \State state $\gets$ Cancelled \Comment{nothing moves}
\EndProcedure
\end{algorithmic}
\end{algorithm}

The two algorithms distinguish the operator's responsibility to submit the daily records and proofs from the finalization right, which any participant may exercise once the market smart contract has verified the day.

\paragraph{The settlement token}
Once the market contract accepts the daily settlement, it must transfer the resulting monetary amounts. It uses an ERC-20 token as a settlement instrument rather than issuing a new currency: members deposit existing tokens into the arrangement and may request withdrawals. The token's native 18-decimal representation interfaces with the protocol's $10^{-12}$-USD accounting unit at the deposit and withdrawal boundary, where the market smart contract requires whole internal units. 

\paragraph{The verifiers}
Two verifier contracts check cryptographic statements before the market updates its state. The batch verifier checks daily settlement calculations, while the smaller reveal verifier supports balance disclosure during recourse and independent exit. The proof-system compiler generates both verifier contracts from the corresponding circuits, keeping their verification logic aligned with the implemented constraints. We introduce those constraints and the difference between the two proofs in Section~\ref{sec:proofs}.

\paragraph{The grid tariff contract}
The pricing rules require reference feed-in and retail tariffs,
which the grid tariff contract supplies independently of the
community operator. 

The contract supports two modes: Under
a predetermined schedule, an administrator publishes the
applicable tariffs, with updates taking effect no earlier
than the following day. Under a variable tariff, $V$
authorized reporters submit daily vectors of 96 tariff
pairs, and the contract accepts the first vector receiving
$q$ identical submissions. This quorum limits unilateral
control over tariff reporting without guaranteeing the
accuracy of the underlying prices. Algorithm~\ref{alg:tariff}
specifies the voting procedure, validation of tariff bounds,
and fallback to the most recent accepted vector.

\begin{algorithm}[H]
\caption{Grid tariff contract}\label{alg:tariff}
\scriptsize\linespread{0.93}\selectfont
\begin{algorithmic}[1]
\Require Mode fixed at deployment. In feed mode, a set of $V$ reporters and a
quorum $q \le V$, also fixed at deployment.
\Ensure The band of each session. In feed mode it is set only by $q$
identical submissions.
\Procedure{submitDailyPrices}{$\mathit{day}, \underline{\boldsymbol\lambda}, \overline{\boldsymbol\lambda}$}
  \State \textbf{require} feed mode, sender is a reporter, $\mathit{day}$ is not past, sender has not reported for $\mathit{day}$
  \For{$t = 1$ \textbf{to} $96$}
    \State \textbf{require} $\underline{\lambda}_t \le \overline{\lambda}_t$
  \EndFor
  \State $\eta \gets \mathrm{keccak256}(\underline{\boldsymbol\lambda}, \overline{\boldsymbol\lambda})$
  \If{$\eta$ not yet stored for $\mathit{day}$}
    \State store $(\underline{\boldsymbol\lambda}, \overline{\boldsymbol\lambda})$ under $\eta$
  \EndIf
  \State $\mathit{votes}[\mathit{day}][\eta] \gets \mathit{votes}[\mathit{day}][\eta] + 1$
  \If{$\mathit{votes}[\mathit{day}][\eta] = q$}
    \State $\mathit{active}[\mathit{day}] \gets \eta$ \Comment{the first hash to reach the quorum wins}
  \EndIf
\EndProcedure
\Procedure{getPrices}{$\tau$}
  \State $\mathit{day} \gets \lfloor \tau / 86400 \rfloor$, \quad $t \gets \lfloor (\tau \bmod 86400)/900 \rfloor$
  \If{schedule mode}
    \State \Return feed-in rate and retail rate of the profile in force, peak or off-peak at $\tau$
  \EndIf
  \If{no hash is active for $\mathit{day}$}
    \State $\mathit{day} \gets$ latest earlier day with an active hash \Comment{stale prices rather than a halt}
  \EndIf
  \State \Return $(\underline{\lambda}_t, \overline{\lambda}_t)$ stored under $\mathit{active}[\mathit{day}]$
\EndProcedure
\end{algorithmic}
\end{algorithm}

\subsection{Off-chain components}\label{sec:offchain}

The protocol requires three off-chain components:

\paragraph{The operator daemon}
The operator runs a session daemon that collects individual
electricity positions, computes aggregate supply and demand,
and submits them to the market contract for price determination.
At the end of each day, it publishes commitments to the
individual positions used in the financial calculations.
A separate data service provides signed receipts and responds
to household information requests. 

These records are essential:
although the ledger verifies the calculations performed on
committed positions, members must also be able to check
that those positions match what the operator originally
acknowledged.

\paragraph{Receipts}
For each fifteen-minute session, the operator issues an
EIP-712 signed receipt~\cite{eip712} containing the day,
session index, member identifier, and electricity supplied
and demanded, in watt-hours. The signature uses the
operator's address registered in the market contract,
allowing each household to independently verify the
authenticity of the acknowledgment. Unlike a daily
statement of payments, these receipts establish a signed
record of the individual quantities entering the market.

At day close, the operator also signs a packet containing
each member's proposed closing balance and the private seed $\nu_n$ used to
construct the daily position commitment
(Section~\ref{sec:proofs}). Using this seed and the 96
session receipts, the household can independently
reconstruct the commitment and compare it with the value
published on-chain. A mismatch reveals that the operator
committed to positions different from those it acknowledged.
The household may also compare the receipts with its own
meter records.

Signed receipts support recourse when the operator
withholds information. A member can file an on-chain request
requiring the missing records to be provided encrypted to
its own key. The request blocks financial settlement until
it is resolved, subject to the deadlines and cancellation
conditions described in Section~\ref{sec:cycle}.

\paragraph{The prosumer client}
The prosumer client automates these verification procedures.
It submits positions, stores receipts, reconstructs daily
commitments, and checks recorded prices and proof outcomes
against the ledger. It also monitors the member's balance
relative to its confidential participation floor. As soon
as a day closes, if the day-close packet is missing or the
reconstructed commitment does not match the published one,
the client files a data request. Once the operator answers,
the client checks the day again against the disclosed
records and, if they still do not verify, requires the
disclosure of the committed balance. The client triggers
finalization only for days whose records it has verified,
once all proofs have been accepted and the recourse window
has closed. Members therefore retain the ability to verify
their accounts, exercise recourse, and complete approved
financial transfers without further cooperation from the
operator.

\subsection{The daily cycle}\label{sec:cycle}

\begin{figure}[htb!]
\centering
\begin{tikzpicture}[font=\scriptsize, x=1.08cm, y=1cm]
  \fill[gray!15] (0,0.1) rectangle (5.6,1.2);
  \node[align=center] at (2.8,0.65) {\textbf{Day $e$: 96 sessions of 15 min}\\aggregates $(s,d)$ posted, prices $r,c$ recorded\\no money moves};
  \fill[gray!15] (5.7,0.1) rectangle (10.3,1.2);
  \node[align=center] at (8.0,0.65) {\textbf{Proof window (12 h, day $e{+}1$)}\\batch proofs submitted, verified,\\then queued};
  \fill[gray!40] (10.4,0.1) rectangle (11.6,1.2);
  \node[align=center] at (11.0,0.65) {\textbf{deadline}};
  \draw[thick] (0,0) -- (11.6,0);
  \foreach \x in {0,5.65,10.35,11.6} {\draw (\x,-0.1) -- (\x,0.1);}
  \node[below] at (0,-0.1) {00:00 ($e$)};
  \node[below] at (5.65,-0.1) {00:00 ($e{+}1$)};
  \node[below] at (10.35,-0.1) {12:00 ($e{+}1$)};
  \draw[thick] (5.65,-0.1) -- (5.65,1.75);
  \node[above, align=center] at (5.65,1.75) {\textbf{midnight}: commitments published, state frozen\\(members, pending deposits, withdrawal requests, floors)};
  \draw[|-|, thick] (5.65,-0.85) -- (10.35,-0.85);
  \node[below, align=center] at (8.0,-0.9) {\textbf{recourse window}\\data request (blocks settlement), clear reveal of a balance,\\cancellation on verifiable grounds only};
  \node[draw, rounded corners=2pt, align=center, inner sep=3pt, fill=gray!8, text width=5.2cm] (ok) at (2.8,-3.65) {\textbf{settled} (anyone may trigger)\\commitments advance, frozen queues consumed,\\capped withdrawals paid, grid leg paid};
  \node[draw, rounded corners=2pt, align=center, inner sep=3pt, fill=gray!8, text width=5.2cm] (ko) at (8.6,-3.65) {\textbf{canceled}\\nothing moves, queues survive,\\the day was traded at the grid tariff};
  \draw[thick] (11.6,0.65) -- (12.0,0.65) -- (12.0,-2.65) -- (8.6,-2.65);
  \draw[-{Stealth[length=1.8mm]}, thick] (8.6,-2.65) -- node[pos=0.5, above, font=\tiny]{all proofs in, no request pending} (2.8,-2.65) -- (ok.north);
  \draw[-{Stealth[length=1.8mm]}, thick] (8.6,-2.65) -- node[right, font=\tiny]{otherwise} (ko.north);
\end{tikzpicture}
\caption{Cycle of one day.}
\label{fig:timeline}
\end{figure}
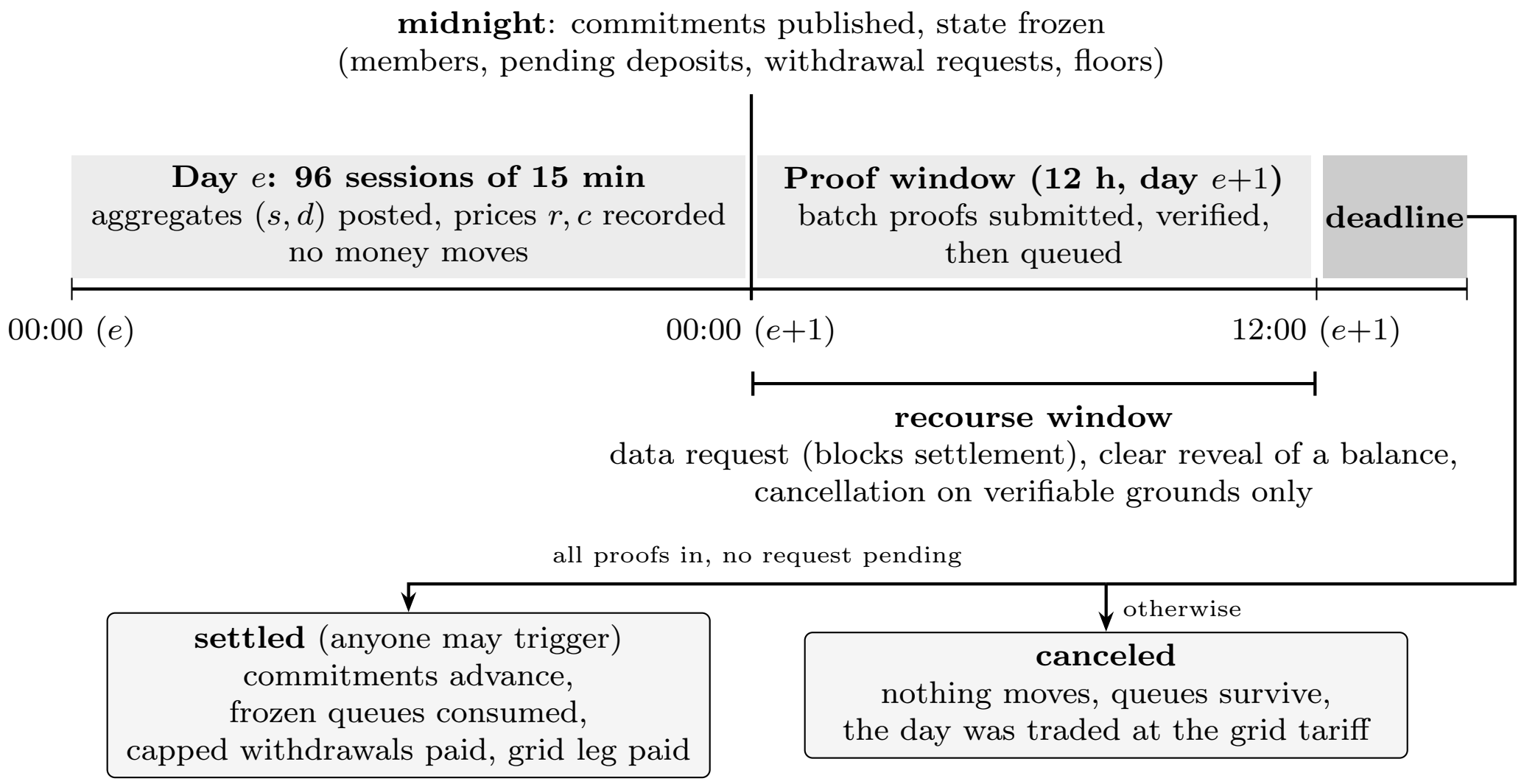

The protocol determines electricity prices every fifteen
minutes but processes financial transfers once per day,
allowing time for cryptographic verification and member
recourse. Figure~\ref{fig:timeline} illustrates the daily
cycle and its possible outcomes.

\paragraph{During the day}
At the end of each fifteen-minute interval, the operator
submits aggregate supply and demand during the following
reporting session. The market contract records these quantities
and computes buyer and seller prices from the applicable grid
tariffs. Individual positions remain off-chain, and no money
changes hands.

\paragraph{At midnight}
The operator publishes commitments to each member's 96
positions and freezes the membership list, deposits, withdrawal
requests, and participation floors. Subsequent changes take
effect the following day, ensuring that the inputs to financial
settlement remain fixed throughout verification.

\paragraph{Verification and recourse}
During the twelve-hour window, the operator submits zero-knowledge proofs in
batches of eight households. Each proof verifies individual payments and
balance updates, while the market contract reconciles the batches' aggregates
and monetary totals with the recorded sessions. All updates remain pending
until these checks succeed. Meanwhile, members have two levels of recourse.
At the first level, a member requests its missing receipts or account
information, which the operator must publish encrypted to the member's key.
Such a request blocks settlement and gives the operator four hours to respond.
If the answer does not let the member verify its day, the member can move to
the second level and require the operator to disclose its committed balance in
the clear, together with a cryptographic proof that this value opens the
commitment recorded on the ledger. This disclosure makes the balance public,
so it is reserved for disputes; it blocks settlement in the same way and is
subject to the same four-hour deadline. Because the prosumer client files the
first request as soon as the day closes, the operator must answer well before
the window ends, which leaves the member time to reach the second level.

\paragraph{Finalization}
Once the verification window closes, anyone may finalize the
day, provided all proofs are accepted, the aggregates reconcile,
and no blocking request remains. Prosumer clients trigger it
only for days whose records they have verified. The contract
then atomically updates account commitments, processes deposits
and capped withdrawals, and settles the community's net grid
position. No partial transfers are possible.

\paragraph{Failure and recovery}
Missing proofs, unanswered requests, or failure to finalize
within the additional six-hour grace period provide grounds
for cancellation. The contract leaves previously settled
balances unchanged and keeps the frozen deposits and withdrawal
requests for the next close; electricity trading for that day
must instead be accounted for through ordinary grid-billing
arrangements. If no day is settled for fourteen days, members
can independently prove and withdraw their last settled
balances, provided no day is closing.

\subsection{Design choices}\label{sec:choices}

Several design choices have important implications and are worth discussing:

\paragraph{Rounding and budget balance}
Seller prices are rounded down and buyer prices up, while the feed-in bound is
rounded up and the retail bound down. Since the exact prices satisfy the budget
identity of Section~\ref{sec:mechanism}, the residual is then non-negative in
every session and under every rule, so the market never pays out more than it
collects. Rounding moves each price by at most a few grains of
$10^{-9}$~USD per kilowatt-hour, so the residual, sent to a reserve, is
negligible.

\paragraph{Participation floors}
A member submits its position before knowing the price it will pay,
since prices depend on the session aggregates.
Instead of locking collateral at every session, each account must hold a
participation floor at the start of the day, covering the most it can owe in
one day: its subscribed power times the highest retail price, over
twenty-four hours. The member thus remains solvent whatever positions it
takes and whatever prices result. A floor set too high, however, would exclude
the member from trading. A change of floor therefore requires two parties: the
operator proposes it and a separate floor administrator confirms it, so that
neither can exclude a member alone. Because the floor would reveal the subscribed power, it is
stored as a commitment and checked inside the proof.

\paragraph{Daily netting}
Payments are settled daily because recourse needs time: a member must be able
to notice a discrepancy, file a request and receive an answer before any money
moves, which cannot fit within a fifteen-minute session. Prices remain fixed
session by session, so the properties of Section~\ref{sec:mechanism} are
unchanged.

\paragraph{Withdrawal capping}
Withdrawals are paid up to the balance available after settlement, a cap
enforced by the circuit, so they cannot break solvency.

\section{Cryptographic guarantees}\label{sec:proofs}

The verification layer relies on two types of cryptographic
proofs. The first establishes that individual payments and
account updates are consistent with committed electricity
positions and the prices recorded on-chain, without revealing
the underlying household data. The second allows a member
to prove the value of a committed balance when recourse or
independent withdrawal requires it. Both are implemented
as Noir circuits~\cite{noir_docs}, with proofs generated
using UltraHonk and the Barretenberg backend over
BN254~\cite{barretenberg}. UltraHonk uses a universal
reference string, avoiding a circuit-specific trusted
setup, while Poseidon2~\cite{poseidon2} provides the
cryptographic commitments.

\subsection{Commitments}
A cryptographic commitment fixes a value without revealing
it. We write $\mathsf{H}(x,\beta)$, where $\beta$ is a fresh
random blinding factor. Randomness matters because account
balances and participation thresholds may otherwise be
recovered by testing values within their limited ranges.
For each member, the market contract stores commitments
$\mathsf{H}(M_n,\beta_n)$ to its balance and
$\mathsf{H}(f_n,\varphi_n)$ to its participation floor.
The latter is jointly established by the operator and a
separate floor administrator. Both values remain private,
while their commitments are updated or frozen according
to the daily accounting cycle.

\paragraph{Committing to a day of positions}
The operator must also commit to the complete sequence of
electricity positions used to calculate payments. For
member $n$, let $(s_{n,t},d_{n,t})$ denote the energy
supplied and demanded in session $t$, measured in
watt-hours. A private seed $\nu_n$ initializes the
following hash chain:
\begin{equation}\label{eq:chain}
\begin{aligned}
\zeta_{-1} &= \mathsf{H}(n,\nu_n),\\
\zeta_t &= \mathsf{H}(\zeta_{t-1},s_{n,t},d_{n,t}),
\qquad t=0,\ldots,95,\\
C_n &= \zeta_{95}.
\end{aligned}
\end{equation}
The resulting commitment $C_n$ binds both the quantities
and their chronological order. The private seed also hides
otherwise predictable sequences, including a day with no
electricity exchange. Crucially, the household can
reconstruct $C_n$ from its 96 signed receipts and the seed
provided in the day-close packet. A mismatch with the
published commitment reveals an inconsistency in the
operator's records, although it does not establish which
reported quantity reflects physical delivery.

\subsection{Batch proofs}\label{sec:batch-proofs}
\paragraph{Batch construction and padding}
We use a fixed circuit size of $Q=8$ members, allowing
communities of $N$ members to be verified through
$J=\lceil N/Q\rceil$ independent batches. Lane $i$ of
batch $j$ corresponds to member slot $jQ+i+1$.

The member count is frozen at midnight; subsequent
registrations enter the next day's accounting. Unused
lanes in the final batch are assigned slot zero, with
zero positions and balances enforced by the circuit.
Their commitments are predefined constants computed
using the same Poseidon2 function, preventing padding
from introducing fictitious transactions.

\begin{figure}[!htbp]
\centering
\newcommand{\bul}[1]{\par\noindent\hangindent=0.8em\hangafter=1\makebox[0.8em][l]{\textbullet}#1}
\begin{tikzpicture}[
    font=\scriptsize, x=1cm, y=1cm,
    box/.style={
        draw, rounded corners=2pt, align=left,
        inner sep=4pt, fill=white, text width=3.3cm,
        execute at begin node={\hyphenpenalty=10000\linespread{1.12}\selectfont}
    },
    pub/.style={box, fill=gray!10},
    flow/.style={-{Stealth[length=1.8mm]}, thick}
]
    \node[box, anchor=north] (priv) at (0,3.2) {
        \textbf{Private inputs} (operator)
        \bul{$(s_{n,t},d_{n,t})$, 96 sessions}
        \bul{seeds $\nu_n$}
        \bul{$M_n$, $f_n$ and their blindings $\beta_n$, $\varphi_n$}\par
    };

    \node[pub, anchor=north] (pubin) at (0,0.55) {
        \textbf{Public inputs} (chain)
        \bul{$C_n$, $\mathsf{H}(M_n,\beta_n)$, $\mathsf{H}(f_n,\varphi_n)$}
        \bul{prices $r_t, c_t$}
        \bul{frozen $\mathrm{dep}_n$, $\mathrm{req}_n$}\par
    };

    \node[box, anchor=north, text width=5.5cm]
        (circ) at (5.18,2.75) {
        \textbf{Batch circuit proves} ($n \in j$, every $t$)
        \bul{private values open their commitments}
        \bul{$\sum_{n\in j} s_{n,t} = s^{(j)}_t$ and $\sum_{n\in j} d_{n,t} = d^{(j)}_t$}
        \bul{$\mathrm{out}_n = \sum_t r_t s_{n,t}$ and $\mathrm{in}_n = \sum_t c_t d_{n,t}$}
        \bul{$M'_n = M_n + \mathrm{out}_n + \mathrm{dep}_n - \mathrm{in}_n - w_n \ge 0$}
        \bul{$\sum_{n\in j} \mathrm{out}_n = \mathrm{out}^{(j)}$, $\sum_{n\in j} \mathrm{in}_n = \mathrm{in}^{(j)}$}
        \bul{$M_n < f_n \Rightarrow s_{n,t} = d_{n,t} = 0$}\par
    };

    \node[pub, anchor=north] (pubout) at (10.36,3.2) {
        \textbf{Public outputs}
        \bul{$s^{(j)}_t$, $d^{(j)}_t$ for every $t$}
        \bul{commitments to $M'_n$}
        \bul{paid withdrawals $w_n$}
        \bul{totals $\mathrm{out}^{(j)}$, $\mathrm{in}^{(j)}$}\par
    };

    \node[pub, anchor=north] (chain) at (10.36,0.85) {
        \textbf{On-chain reconciliation}
        \bul{$\sum_j s^{(j)}_t = s_t$}
        \bul{$\sum_j d^{(j)}_t = d_t$}
        \bul{$\sum_j \mathrm{out}^{(j)} = \sum_t r_t s_t$}
        \bul{$\sum_j \mathrm{in}^{(j)} = \sum_t c_t d_t$}\par
    };

    \draw[flow] (priv.east) -- ($(circ.west)+(0,0.6)$);
    \draw[flow] (pubin.east) -- ($(circ.west)+(0,-0.6)$);
    \draw[flow] ($(circ.east)+(0,0.6)$) -- (pubout.west);
    \draw[flow] (pubout.south) -- (chain.north);
\end{tikzpicture}
\caption{Interface of the batch circuit.}
\label{fig:proof}
\end{figure}
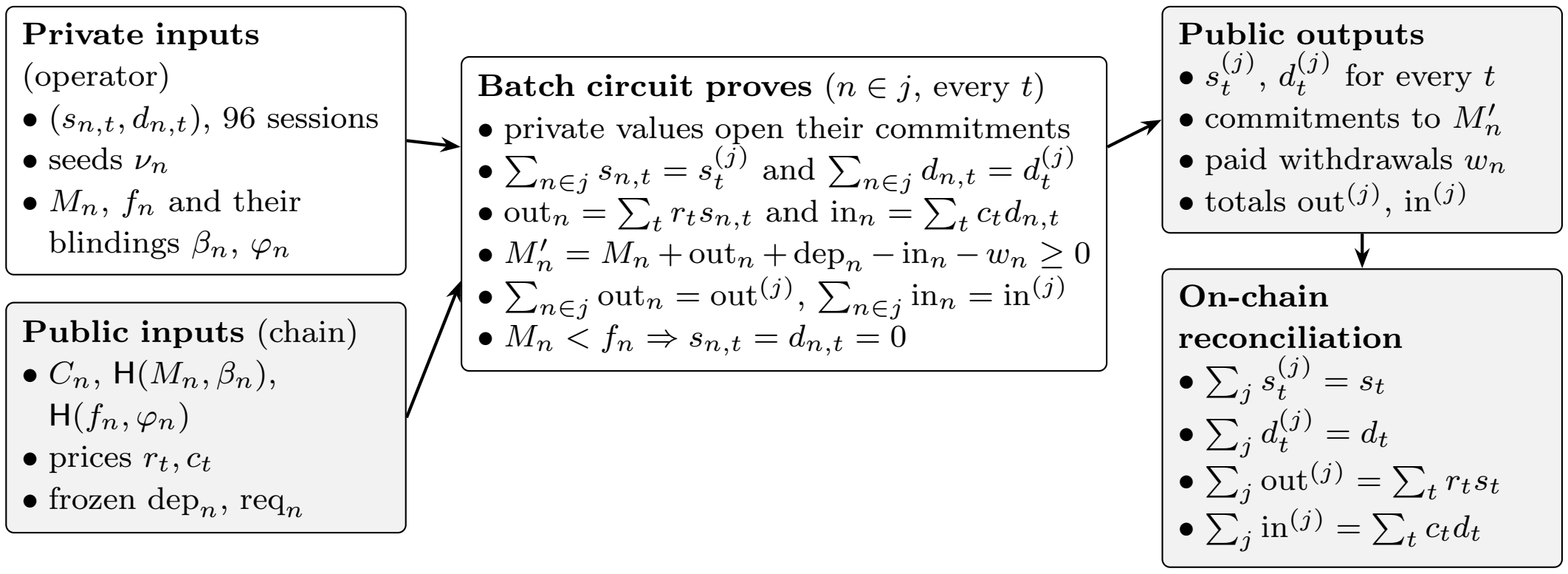

\paragraph{Public inputs and private witnesses}
Figure~\ref{fig:proof} summarizes the batch circuit. The market contract assembles its 450 public field
elements from recorded on-chain values and the outputs
submitted with the proof. These include the 96 buyer
and seller prices, member identifiers, position, balance
and floor commitments, frozen deposits and withdrawal
requests, proposed balance updates, and batch-level
energy and monetary totals. In particular, the operator
cannot substitute arbitrary prices or opening
commitments.

The private inputs contain the $8\times96\times2$
electricity quantities, opening balances, participation
floors, commitment blinding factors, and private seeds.
The circuit verifies relations between these private
values and the public records without disclosing the
underlying household data.

The batch circuit enforces six conditions connecting
committed electricity positions to public prices and
individual accounts.

\paragraph{Binding and aggregation}
First, the private positions, opening balances, and
participation floors must reproduce their respective
commitments. The operator therefore cannot choose
different private values when constructing the proof.
Second, for every session $t$, the circuit aggregates
the positions of the members in batch $j$ and requires
\begin{equation}
\sum_{n\in j}s_{n,t}=s_t^{(j)},\qquad
\sum_{n\in j}d_{n,t}=d_t^{(j)}.
\end{equation}
These quantities represent the batch's contribution to
the community's total supply and demand.

\paragraph{Payments and balance updates}
Third, the circuit calculates each member's monetary
inflows and outflows using the prices previously
recorded on-chain:
\begin{equation}\label{eq:amounts}
\mathrm{out}_n=\sum_{t=0}^{95}r_t s_{n,t},
\qquad
\mathrm{in}_n=\sum_{t=0}^{95}c_t d_{n,t}.
\end{equation}
This establishes the link between public price
determination and private financial accounting.
The circuit applies the recorded prices rather than
recomputing the pricing rule, so the same proof
verifies all three mechanisms. Positions are recorded
in watt-hours and prices at a precision of $10^{-9}$
USD/kWh, yielding exact integer monetary amounts
in units of $10^{-12}$ USD.

Fourth, the proof checks the resulting account balance:
\begin{equation}
\begin{aligned}
M'_n &= M_n+\mathrm{out}_n+\mathrm{dep}_n
       -\mathrm{in}_n-w_n,\\
w_n &= \min\{\mathrm{req}_n,\,
       M_n+\mathrm{out}_n+\mathrm{dep}_n-\mathrm{in}_n\}.
\end{aligned}
\end{equation}
Here $\mathrm{dep}_n$ and $\mathrm{req}_n$ are the
deposit and withdrawal request frozen at midnight.
Withdrawals cannot exceed the available balance.

Since circuit arithmetic takes place over a finite
field, unsigned 64-bit range checks on intermediate
and final balances also prevent negative amounts
from wrapping around into apparently valid positive
values. Energy quantities are similarly constrained
to 32 bits, matching their on-chain representation.

\paragraph{Conservation and participation}
Fifth, the circuit requires the members' proven
inflows and outflows to sum to the batch totals:
\[
\sum_{n\in j}\mathrm{in}_n=\mathrm{in}^{(j)},
\qquad
\sum_{n\in j}\mathrm{out}_n=\mathrm{out}^{(j)}.
\]
This prevents discrepancies between individual
payments and the monetary amounts declared for
the batch.

Finally, the circuit enforces participation eligibility
without disclosing account balances or floors.
A member whose opening balance is below its
participation floor must have zero positions
throughout the day:
\[
M_n<f_n
\quad\Longrightarrow\quad
s_{n,t}=d_{n,t}=0
\qquad \forall t.
\]
The circuit verifies this condition against the
committed values and also requires
$s_{n,t}d_{n,t}=0$, ruling out simultaneous buying
and selling within a session. Thus, ineligible
members cannot contribute electricity to the
verified aggregates, while their financial
thresholds remain confidential.

\subsection{On-chain reconciliation}
A valid batch proof establishes the consistency
of its members' accounts, but does not by itself
verify the community aggregates used to determine
market prices. The market contract therefore
combines the partial quantities from all $J$
verified batches and requires, for every session,
\begin{equation}\label{eq:reconcile}
\sum_j s_t^{(j)}=s_t,\qquad
\sum_j d_t^{(j)}=d_t.
\end{equation}
It also reconciles monetary totals with the
recorded prices:
\[
\sum_j\mathrm{out}^{(j)}=\sum_t r_t s_t,
\qquad
\sum_j\mathrm{in}^{(j)}=\sum_t c_t d_t.
\]
Finally, the contract checks that the residual,
including the grid exchange and conservative
rounding, is non-negative. Only after these
conditions are satisfied does it apply the new
balance commitments, process withdrawals, and
transfer the amounts owed to or by the grid
account.

Together, the batch proofs and on-chain
reconciliation establish the central verification
guarantee: individual payments are calculated
from committed positions using prices determined
by the same community aggregates. The operator
cannot independently alter the quantities, payments,
or balance updates while obtaining an accepted
financial settlement. Members can also detect
commitments inconsistent with their signed receipts.
The guarantee remains conditional on the reported
inputs: an incorrect position consistently
acknowledged and committed by the operator may
still pass these cryptographic checks.

\subsection{Balance disclosure and independent withdrawal}

The batch proofs verify account updates without revealing
their underlying balances. Recourse and independent withdrawal,
however, sometimes require a member to establish the value
hidden by an existing commitment. For this purpose, a second,
smaller circuit verifies that a disclosed balance and its
private blinding factor reproduce the recorded commitment,
subject to the same 64-bit range constraint.

\subsection{Limits of the guarantees}\label{sec:residual}

\paragraph{Reported positions}
No proof can establish that committed positions match what households actually
sent or consumed~\cite{chiu2024}. An operator that signs and commits to the
same wrong figure passes every check, and the dispute must be resolved outside
the protocol, for instance against the load curves held by the distribution
system operator.

\paragraph{Confidentiality towards the prover}
The prover sees positions in clear. The community operator must obtain these data through an authorized metering-data arrangement, subject to local requirements.\footnote{In French collective self-consumption, for instance, the organizing entity already receives each participant's load curve from the distribution system operator~\cite{enedis_acc}.}

\paragraph{Scalability and admission}
Each eight-member batch adds one proof, generated in parallel and verified at
constant cost, so costs grow linearly with community size. Who may join a
community is left to its governance.

\section{Experimental Testing}\label{sec:sim}

We now test the functioning of the proposed protocol with two experiments: First, we assess the performance of the settlement infrastructure, in particular its execution and verification features, assuming that households retain their
recorded electricity positions.  Then, we test the economics of the protocol, which includes assessing the gains from local trade and the incentives induced by the mechanism under strategic scheduling. 

\subsection{Data and experiment design}\label{sec:data}\label{sec:protocol}

We construct a ten-household community using the US National Renewable Energy Laboratory's ResStock end-use profiles~\cite{nrel_eulp}, which simulate consumption and rooftop photovoltaic generation at
fifteen-minute resolution. We select ten dwellings in one Massachusetts
county, five with photovoltaic generation and five without. This composition
ensures the presence of both net sellers and buyers. The experiment covers four
consecutive summer days, 1--4 July 2018, for which the load and generation
profiles coincide with the dates of the price observations.

We construct the lower tariff bound from the corresponding ISO New England
hourly day-ahead locational marginal price~\cite{isone_smd}, and set the retail
bound 12 US cents/kWh higher. Thus, if
$\lambda_h$ denotes the wholesale price in hour $h$, the two bounds for every
fifteen-minute interval within that hour are
$\lamlow_t=\lambda_h$ and $\lamhigh_t=\lambda_h+0.12$ USD/kWh.

For each quarter-hour, household net supply is its photovoltaic generation
less consumption. We do not introduce batteries or flexible-load responses
in this first experiment. We execute a complete four-day cycle separately for
each pricing rule, retaining exactly the same physical quantities and tariff
bounds. 

\subsection{Technical testing Execution, verification and ledger availability}\label{sec:perf}\label{sec:besu}

We first examine the protocol's computational requirements and the
consistency of the resulting balances. Table~\ref{tab:perf} reports the on-chain
resource consumption for one full day with ten households. The protocol
opens 96 sessions and verifies two eight-lane proof batches; the latter
account for most of the daily gas. The complete cycle uses approximately
19.0 million gas and 46 kB of transaction data. Generating an eight-member
proof takes approximately 2.4 seconds on one thread. 

\begin{table}[!htbp]
\centering
\caption{Daily on-chain execution for ten households (mid-market rate).
Measurements are averaged over the four-day run; amounts are rounded.}
\label{tab:perf}
\footnotesize\setlength{\tabcolsep}{5pt}
\begin{tabular}{@{}p{5.0cm}rrr@{}}
\toprule
Operation & Calls/day & Gas/day & Calldata/day \\
\midrule
Record aggregates and prices & 96 & 6.3 M & 12.7 kB \\
Post position commitments & 1 & 0.7 M & 0.4 kB \\
Submit and verify batch proofs & 2 & 11.4 M & 32.8 kB \\
Finalize daily settlement & 1 & 0.7 M & $<$0.1 kB \\
\midrule
\textbf{Total} & & \textbf{19.0 M} & \textbf{46 kB} \\
\bottomrule
\end{tabular}
\end{table}

The computation scales primarily with the number of proof batches, rather
than the number of session-price calculations. At eight members per batch,
a 200-member community would require 25 proofs and approximately 140 million
gas per day. The three pricing rules
have similar resource requirements. 

We independently recomputed each member's closing balance from its signed
session receipts, the prices recorded on-chain, and its deposits and paid
withdrawals. All 40 member-days matched the operator's signed closing
records under each pricing rule, down to the internal accounting unit.
Conservative rounding also produced a positive but negligible reserve
residual---less than $6\times10^{-7}$ USD over the four days under any rule.
The tests exercised the initial member data-request procedure once each day.

We next test the ledger's ability to meet the protocol's timing
and availability requirements on our four-validator Hyperledger
Besu network (Table~\ref{tab:besu-robustness}). Under normal
operation, transactions reach finality in approximately five
seconds, well within the fifteen-minute reporting interval.
The network continues producing blocks with one validator
offline, but halts when two validators become unavailable,
preventing the formation of a QBFT quorum. Restoring the quorum
preserves the agreed block history, although accumulated
consensus timeouts delay the resumption of block production.
Similarly, clock offsets primarily reduce the affected
validator's participation without disrupting consensus among
correctly synchronized nodes. These results illustrate the
protocol's fault-tolerance properties under the tested
conditions.

\begin{table}[!htbp]
\centering
\caption{Fault tolerance and clock synchronization.}
\label{tab:besu-robustness}
\footnotesize
\setlength{\tabcolsep}{4pt}
\renewcommand{\arraystretch}{1.08}
\begin{tabular}{@{}llrl@{}}
\toprule
Experiment & Condition & Block interval (s) & Outcome \\
\midrule
\multicolumn{4}{@{}l}{\textit{Panel A. Validator availability}} \\
Baseline & 4/4 online & 5.1 & Operational \\
One failure & 3/4 online & 8.1 & Operational \\
Two failures & 2/4 online & -- & Consensus halted \\
Recovery & Quorum restored & -- & Delayed resumption \\
\addlinespace[4pt]
\multicolumn{4}{@{}l}{\textit{Panel B. Validator clock skew}} \\
Baseline & No offset & 5.0 & Normal operation \\
Clock ahead & $+2$ s & 8.2 & Proposals rejected \\
Clock behind & $-3$ s & 7.7 & Reduced participation \\
Clock behind & $-8$ s & 8.0 & No proposals accepted \\
Clock behind & $-30$ s & 8.2 & No proposals accepted \\
\bottomrule
\end{tabular}

\vspace{2pt}
\begin{minipage}{0.98\linewidth}
\scriptsize
\textit{Notes:} Mean block intervals are measured during
two-minute validator-failure phases and four-minute
clock-skew phases. After quorum restoration, block
production resumed with a delay of 5 min 14 s.
All validators retained the same block history.
Tests ran on four containers hosted on a single machine.
\end{minipage}
\end{table}

\subsection{Local matching and the allocation of gains}\label{sec:results} 

\begin{figure}[!htbp]
\centering
\includegraphics[width=0.7\textwidth]{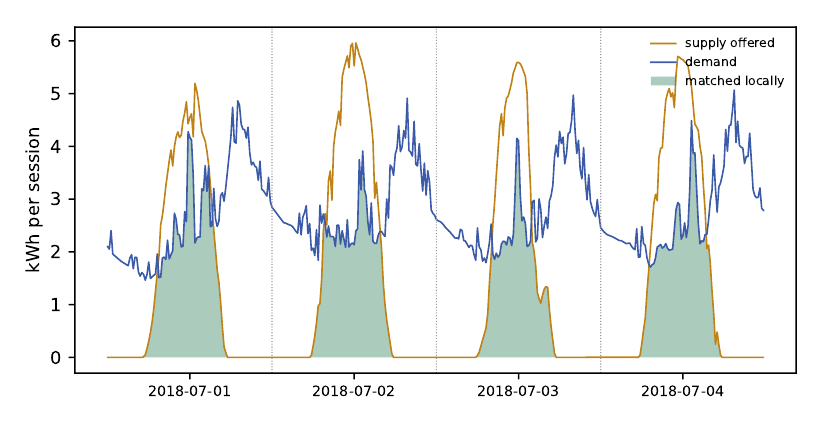}
\caption{Aggregate supply, demand and locally matched electricity in each
fifteen-minute interval, 1--4 July 2018.}
\label{fig:raw-sessions}
\end{figure}

Having established the accuracy of executed settlements in the test
environment, we examine their implied economic efficiency gains. For fixed household
positions, the three pricing rules match the same 366.0 kWh of electricity
over four days: 34.4\% of total demand and 61.6\% of total supply.
Figure~\ref{fig:raw-sessions} shows the quarter-hourly supply, demand and
amount matched within the community for settlement. Under the assumed
12-cent tariff spread, matching reduces the community's net grid-only bill
from \$144.01 by \$43.92, or 30.5\%. This is the budget identity in
Section~\ref{sec:mechanism}: with fixed positions, the community gain equals
the tariff spread multiplied by matched electricity. A 6-cent spread would
yield \$21.96, while a 17-cent spread would yield \$62.22; the physical
matching volume would remain unchanged.

\begin{figure}[H]
\centering
\includegraphics[width=.75\linewidth]{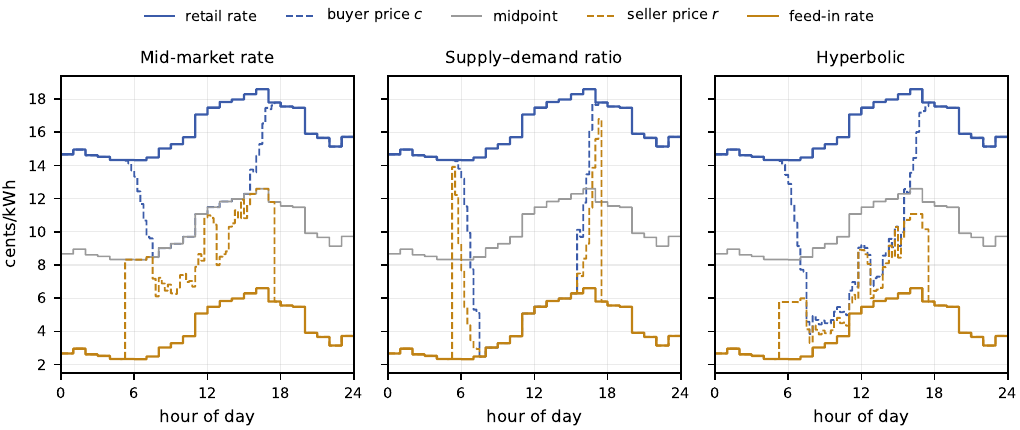}
\caption{Buyer and seller prices under the three pricing rules
on 1 July 2018\addition{, recorded load curves}.}
\label{fig:raw-prices}
\end{figure}

The pricing rule determines how households share this common gain (Table~\ref{tab:raw-split}). The mid-market rate divides it
equally between buyers and sellers. Supply--demand ratio pricing assigns
almost the entire gain to buyers in this sample, while hyperbolic pricing
produces an intermediate split. Figure~\ref{fig:raw-prices} relates these
differences to the resulting session prices. Every household benefits
relative to grid-only trading at its recorded positions because the average
buying and selling prices remain within the applicable grid tariff bounds.

\begin{table}[H]
\centering
\caption{Allocation of the same \$43.92 fixed-profile gain across pricing
rules. Prices are volume-weighted averages (US cents/kWh); the average
feed-in and retail bounds are 5.57 and 16.64 cents/kWh.}
\label{tab:raw-split}
\small\setlength{\tabcolsep}{6pt}
\begin{tabular}{@{}lrrrr@{}}
\toprule
 & \multicolumn{2}{c}{Gain (USD)} & \multicolumn{2}{c}{Mean price (c/kWh)} \\
\cmidrule(lr){2-3}\cmidrule(l){4-5}
Rule & Sellers & Buyers & Sellers receive & Buyers pay \\
\midrule
Mid-market rate & 21.96 & 21.96 & 9.27 & 14.58 \\
Supply--demand ratio & 1.75 & 42.17 & 5.87 & 12.68 \\
Hyperbolic & 8.91 & 35.01 & 7.07 & 13.35 \\
\bottomrule
\end{tabular}
\end{table}

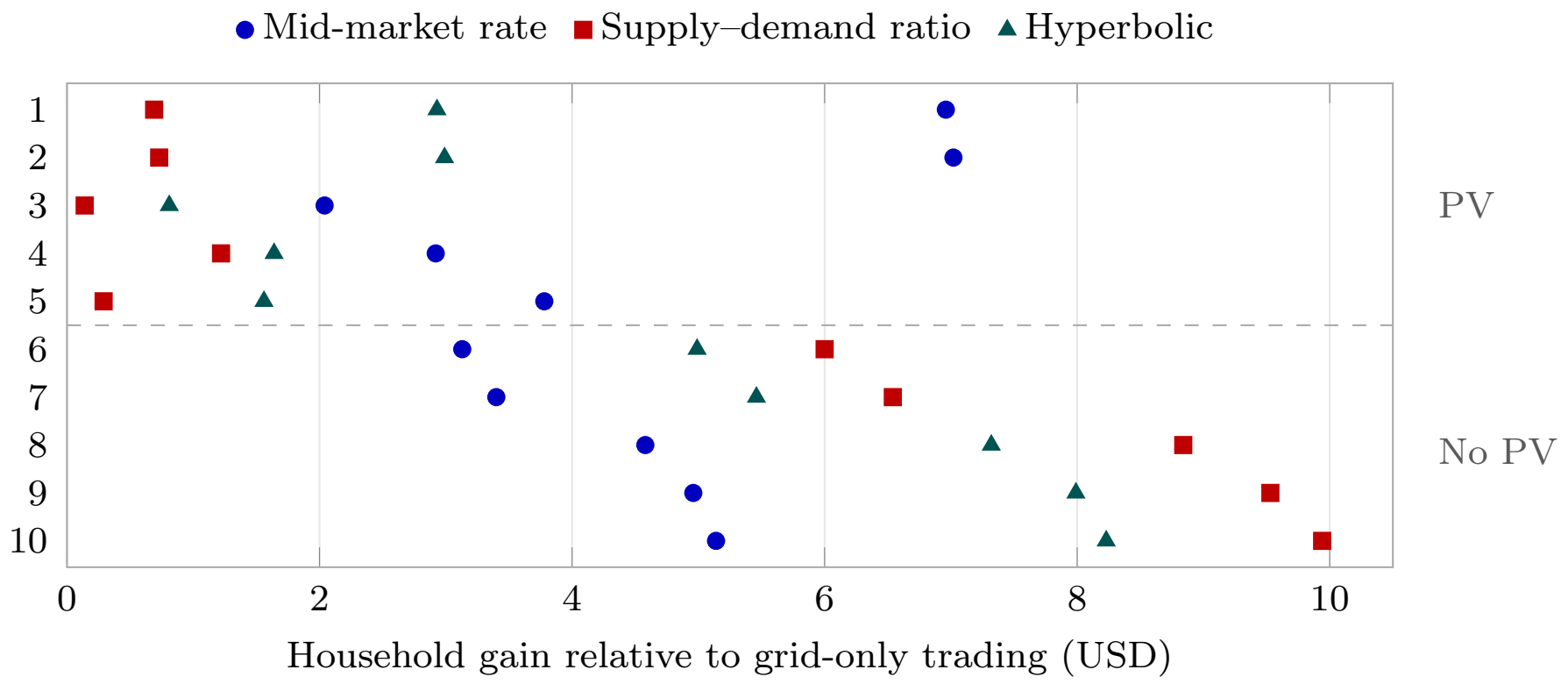
\begin{figure}[H]
\centering
\begin{tikzpicture}
\begin{axis}[
    width=.73\linewidth,
    height=3.65cm,
    scale only axis,
    xmin=0, xmax=10.5,
    ymin=.45, ymax=10.55,
    y dir=reverse,
    xtick={0,2,4,6,8,10},
    ytick={1,2,3,4,5,6,7,8,9,10},
    tick label style={font=\scriptsize},
    ytick style={draw=none},
    xlabel={Household gain relative to grid-only trading (USD)},
    xlabel style={
        font=\scriptsize,
        at={(axis description cs:.5,-.11)},
        anchor=north
    },
    xmajorgrids=true,
    ymajorgrids=false,
    major grid style={gray!19},
    axis line style={gray!65},
    legend columns=3,
    legend style={
        draw=none,
        fill=none,
        at={(.5,1.035)},
        anchor=south,
        font=\scriptsize,
        /tikz/every even column/.append style={column sep=4pt}
    },
    clip=false,
    enlargelimits=false
]
\addplot[only marks,mark=*,mark size=1.75pt,blue!75!black] coordinates {
    (6.96,1) (7.02,2) (2.04,3) (2.92,4) (3.78,5)
    (3.13,6) (3.40,7) (4.58,8) (4.96,9) (5.14,10)
};
\addlegendentry{Mid-market rate}
\addplot[only marks,mark=square*,mark size=1.75pt,red!75!black] coordinates {
    (.69,1) (.73,2) (.14,3) (1.22,4) (.29,5)
    (6.00,6) (6.54,7) (8.84,8) (9.53,9) (9.94,10)
};
\addlegendentry{Supply--demand ratio}
\addplot[only marks,mark=triangle*,mark size=2.0pt,teal!65!black] coordinates {
    (2.93,1) (2.99,2) (.81,3) (1.64,4) (1.56,5)
    (4.99,6) (5.46,7) (7.32,8) (7.99,9) (8.23,10)
};
\addlegendentry{Hyperbolic}

\draw[gray!65,dashed,thin]
    (axis cs:0,5.5) -- (axis cs:10.5,5.5);

\node[anchor=west,font=\scriptsize,gray!70!black]
    at (axis description cs:1.02,.75) {PV};

\node[anchor=west,font=\scriptsize,gray!70!black]
    at (axis description cs:1.02,.24) {No PV};

\end{axis}
\end{tikzpicture}
\caption{Household gains under the three pricing rules\addition{, recorded load curves}.}
\label{fig:raw-household-gains}
\end{figure}

The aggregate comparison also conceals substantial differences across
households. Figure~\ref{fig:raw-household-gains} shows the gain of each
household under the three pricing rules. The two largest solar producers
gain most under the mid-market rate, whereas households without
photovoltaic generation gain most under supply--demand ratio pricing.
The hyperbolic rule lies between these outcomes.

\subsection{Pricing and incentives}\label{sec:equilibrium}

The preceding experiment holds household electricity positions fixed. It
therefore establishes the operation of the settlement infrastructure, but not
how households respond to its prices. In a small community, an individual
trade also affects aggregate supply and demand.\footnote{If the community is sufficiently large the price impact becomes negligible. See
 ~\cite{fabi2025}.} We now let households schedule storage and
flexible consumption strategically and compare the resulting equilibria
with centrally coordinated scheduling.

\paragraph{Household decisions}\label{sec:eq-households}\label{sec:eq-model}
We retain the ten households, four days and tariff band that we introduced before. Each household receives a lossless 20~kWh battery,
with a 5~kW charge/discharge limit and a 5~kW grid connection. Some appliances produce inflexible loads (such as cooling,
heating, cooking, lighting, refrigeration and plug loads).
Water heating and pool or spa pumps may shift within the day, preserving
their total daily energy and maximum power. The 40 recorded dishwasher,
washer and dryer cycles may shift their starting times within each day,
without altering their profiles or the order of washing and drying.
These choices deliberately provide favorable flexibility rather than
reproduce the equipment owned by the sampled households.

A schedule $a_n\in\mathcal A_n$ specifies a household's consumption,
battery operation and net market positions over $T=384$ sessions of
$\Delta=0.25$~h. Writing $\omega_{n,t}$ for photovoltaic generation,
$p_{n,t}$ for consumption after rescheduling, $k_{n,t}$ for battery charging
(negative for discharge), $b_{n,t}$ for stored energy and $x_{n,t}$ for net
sales (negative for purchases), its main physical constraints are
\begin{equation}\label{eq:household}
\begin{gathered}
 x_{n,t}=\omega_{n,t}-p_{n,t}-k_{n,t},\quad
 b_{n,t}=b_{n,t-1}+k_{n,t},\\[-3pt]
 0\le b_{n,t}\le20,\quad |k_{n,t}|,|x_{n,t}|\le5\Delta,
 \quad b_{n,0}=b_{n,T}=0.
\end{gathered}
\end{equation}
All quantities above are expressed in energy per session in kWh. The feasible set
$\mathcal A_n$ also enforces the flexible-load and appliance-cycle conditions
stated above, and excludes photovoltaic curtailment.

\label{sec:eq-benchmarks}\paragraph{Benchmarks} To evaluate the gains from trade with optimizing prosumers, we use two welfare benchmarks: each household optimizing alone under the
grid tariff, and a social planner minimizing the community's net grid bill.
With $z^+=\max\{z,0\}$, $z^-=\max\{-z,0\}$ and
$X_t=\sum_n x_{n,t}$, their objectives are
\begin{equation}\label{eq:planner}
\begin{aligned}
 G_n^*&=\min_{a_n\in\mathcal A_n}\sum_t
 (\lamhigh_t x_{n,t}^- -\lamlow_t x_{n,t}^+),\\[-1pt]
 C^{\rm SP}&=\min_{a\in\prod_n\mathcal A_n}\sum_t
 (\lamhigh_t X_t^- -\lamlow_t X_t^+).
\end{aligned}
\end{equation}
The first benchmark accounts for savings achievable through storage and
load shifting even without local exchange; the second also internalizes
the gains from matching households' positions. Under the AMM, household
$n$ instead minimizes
\begin{equation}\label{eq:nash-bill}
 P_n(a_n,a_{-n})=\sum_t
 \bigl(c_t\,x_{n,t}^- -r_t\,x_{n,t}^+\bigr),
\end{equation}
where the mechanism prices $(r_t,c_t)$ depend on the positions of all
households, including $n$. We call a profile $a^*$ an
$\varepsilon$-Nash equilibrium when
\begin{equation}\label{eq:nash}
 P_n(a_n^*,a_{-n}^*)\le P_n(a_n,a_{-n}^*)+\varepsilon
 \quad \forall n,\ a_n\in\mathcal A_n.
\end{equation}
Because average transaction prices remain inside the tariff band,
a household can always replay its grid-only schedule inside the community and
pay at most $G_n^*$; equilibrium participation is therefore individually
rational up to $\varepsilon$. The social planner's allocation does not have
this property: it minimizes the community's total bill and may leave some
households worse off than trading alone.

\paragraph{Numerical experiments and outcomes}
\label{sec:eq-computation}\label{sec:eq-results}

We solve the grid-only and social-planner problems with
Gurobi. For each pricing rule, we compute pure-strategy
equilibrium profiles through sequential best-responses (Algorithm~\ref{alg:best-response}).
Starting from baseline schedules, households take turns
optimizing their entire four-day schedule against the
current schedules of the others. Each household thus
accounts for the effect of its electricity positions on
market prices. We repeat the procedure until no household
has a profitable deviation above the prescribed tolerance,
or for at most 60 passes.

\begin{algorithm}[H]
\caption{Sequential best-response computation}
\label{alg:best-response}
\footnotesize
\begin{algorithmic}[1]
\Require Pricing rule $R$, initial schedules $a^0$,
tolerance $\kappa=10^{-4}$ USD, maximum $K=60$ passes.
\State $a \gets a^0$
\For{$k=1,\ldots,K$}
    \State $\mathrm{updated}\gets\mathrm{false}$
    \For{each household $n$}
        \State $\hat a_n\gets
        \arg\min_{a_n\in\mathcal A_n}P_n(a_n,a_{-n})$
        \If{$P_n(a_n,a_{-n})-
             P_n(\hat a_n,a_{-n})>\kappa$}
            \State $a_n\gets\hat a_n$;
            $\mathrm{updated}\gets\mathrm{true}$
        \EndIf
    \EndFor
    \If{$\mathrm{updated}=\mathrm{false}$}
        \State \textbf{break}
    \EndIf
\EndFor
\State Evaluate the maximum unilateral improvement
$\varepsilon$ at the final profile.
\State \Return $(a,\varepsilon)$
\end{algorithmic}
\end{algorithm}

Optimizing separately against the grid reduces the four-day community bill
from \$144.01 to \$100.22; the planner lowers it further to \$68.03.
Thus, coordination creates \$32.18 of additional gains beyond
individually optimized grid trading. As Table~\ref{tab:eq-community} shows,
the three pricing rules capture 84--85\% of that surplus at the reported
Nash profiles, with little variation in aggregate costs across the rules.

\begin{table}[!htbp]
\centering
\caption{Community outcomes over four days. Gains are relative to
individually optimized grid trading; the percentage is the share of the
planner's additional gain. $\varepsilon$ is the largest gain any household
could still obtain by deviating.}
\label{tab:eq-community}
\footnotesize\setlength{\tabcolsep}{3pt}
\begin{tabular}{@{}lrrrrr@{}}
\toprule
Outcome & Bill (USD) & Gain (USD) & Share (\%) & Local (kWh) & $\varepsilon$ (USD)\\
\midrule
Recorded positions, grid only & 144.01 & -- & -- & -- & --\\
Optimized individually, grid only & 100.22 & 0 & 0 & 0 & --\\
Nash, mid-market rate & 72.88 & 27.34 & 85 & 365.2 & 0.00008\\
Nash, supply--demand ratio & 72.96 & 27.26 & 85 & 377.0 & 0.00100\\
Nash, hyperbolic & 73.33 & 26.89 & 84 & 372.3 & 0.00504\\
Social planner & 68.03 & 32.18 & 100 & 536.6 & --\\
\bottomrule
\end{tabular}
\end{table}

\begin{figure}[!htbp]
\centering
\includegraphics[width=\linewidth]{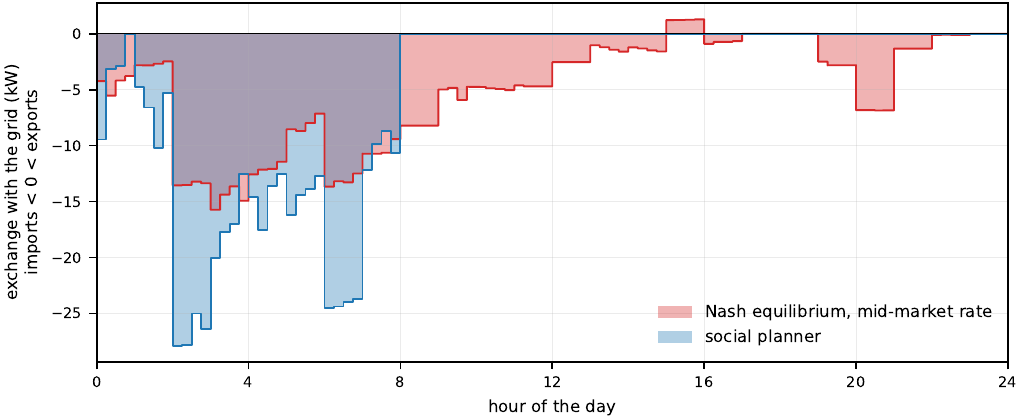}
\caption{Mean exchange of the community with the grid by quarter hour, at the mid-market Nash equilibrium and for the social planner. Imports are negative.}
\label{fig:planner-nash}
\end{figure}

The planner matches 537~kWh locally, compared with approximately
365--377~kWh at equilibrium. This gap reflects two possible incentives:
each household's trade changes its own average price, while that price may
also differ from the value of electricity to the community. For instance,
when the community needs imports, a local sale displaces electricity at the
retail rate, although the seller receives less and may prefer to store
energy for later. The simulation does not identify the relative importance
of these two effects. Figure~\ref{fig:planner-nash} shows their combined
implication: the planner concentrates grid imports in cheaper early hours,
whereas the Nash community continues to buy through much of the day.

\begin{figure}[!htbp]
\centering\includegraphics[width=1\linewidth]{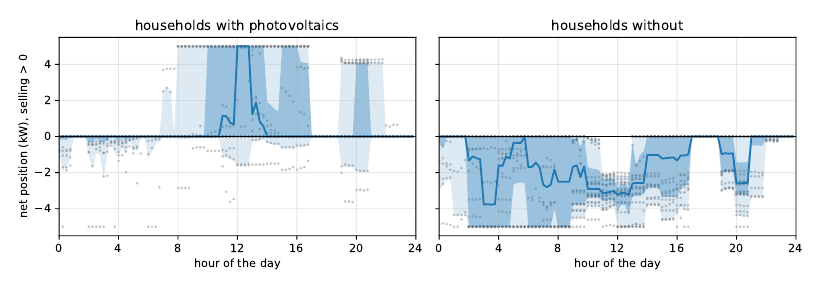}
\centering\includegraphics[width=1\linewidth]{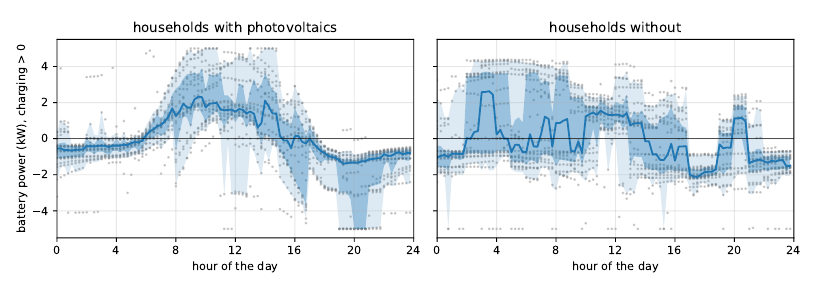}
\caption{Above: Net market positions at the mid-market Nash equilibrium. \\Below: Battery power at the mid-market Nash equilibrium.}
\label{fig:eq-battery-loads}
\end{figure}

\begin{figure}[!htbp]
\centering\makebox[\linewidth][c]{\includegraphics[width=1.4\linewidth]{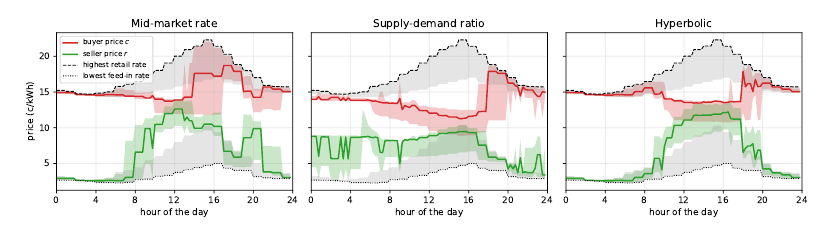}}
\caption{Hourly price distributions at the Nash profiles under the three
pricing rules.}
\label{fig:eq-prices}
\end{figure}

The remaining figures trace the community behavior under the simulated Nash equilibrium. Solar-equipped
households charge batteries during the day and discharge in the evening,
while households without photovoltaics charge more at night. Their trading positions follow this
pattern: producers sell mainly around midday, whereas non-producers
import heavily overnight. Flexible loads and
appliance cycles add a smaller midday shift in consumption
(Figure~\ref{fig:eq-battery-loads}). Endogenous prices reflect the resulting
supply--demand balance: the mid-market seller rate rises during the day,
the supply--demand ratio rule favors buyers, and the hyperbolic rule
lies between them (Figure~\ref{fig:eq-prices}).

Distribution differences for different pricing rules remain important. The two largest solar producers each gain
around USD~7.2 under the mid-market rule, but only USD~2.4 under the
supply--demand ratio rule, which favors consumers; hyperbolic pricing is
intermediate. All households gain at the Nash profiles, whereas the
planner's allocation, valued at the rules' prices, leaves some worse off
than independently optimized grid trading.

\paragraph{Equilibrium selection and robustness}
\label{sec:eq-robustness}

The preceding results concern one equilibrium profile for each
pricing rule. However, the finite-player game can admit multiple
equilibria, potentially associated with different community
welfare. We therefore examine whether the efficiency results
depend on the equilibrium selected by the numerical procedure.

For each rule, we repeat the best-response computation from
20 initial configurations, including the independently optimized
grid schedules, the planner's allocation, random combinations of
the two, and different orders of play. Table~\ref{tab:eq-multistart}
summarizes the results. The procedure terminates before the
60-pass limit in 18 runs under the mid-market rule and 14 runs
under each of the other two rules. The remaining profiles exhibit
maximum measured unilateral improvements of at most
USD~0.0202 over four days.

Although the procedure identifies distinct outcomes, aggregate
welfare varies little. Community bills differ by less than one
dollar within each pricing rule, and the computed profiles
recover between 82.7\% and 88.0\% of the additional surplus
available to the social planner. We obtain similar results when
perturbing the computed profiles by resetting one or three
households to their independently optimized schedules:
individual schedules can change substantially, but community
bills change by at most USD~0.145.

\begin{table}[!htbp]
\centering
\caption{Twenty starting configurations per rule. The worst ratio is the
highest \emph{observed} community bill divided by the planner bill
(USD~68.03), not a global price-of-anarchy estimate.}
\label{tab:eq-multistart}
\footnotesize\setlength{\tabcolsep}{4pt}
\begin{tabular}{@{}lcrrr@{}}
\toprule
Rule & Early stops & Bill range (USD) & Gain (\%) & Worst ratio\\
\midrule
Mid-market rate & 18/20 & 71.90--72.88 & 84.9--88.0 & 1.071\\
Supply--demand ratio & 14/20 & 72.94--73.13 & 84.2--84.8 & 1.075\\
Hyperbolic & 14/20 & 73.26--73.60 & 82.7--83.8 & 1.082\\
\bottomrule
\end{tabular}
\end{table}

\paragraph{Price of Anarchy}
Finally, we assess the quality of matching under decentralization by comparing the worst
equilibria identified numerically with the social planner. 
Across the three pricing rules, the observed cost ratios
range from 1.071 to 1.082 (Table~\ref{tab:eq-multistart}).
For comparison, tariff-bounded prices guarantee that, at
equilibrium, no household pays more than its optimized
grid-only bill. This yields the theoretical upper bound
\begin{equation}\label{eq:poa-bound}
\mathrm{PoA}
\leq \frac{\sum_n G_n^*}{C^{\rm SP}}
=\frac{100.22}{68.03}
\simeq 1.47.
\end{equation}
The numerical results are therefore considerably tighter:
across the equilibria identified, strategic scheduling
generates only a modest efficiency loss relative to
centralized coordination.

\section{Conclusion}\label{sec:conclusion}
We have implemented an automated market maker for local electricity exchange in which fifteen-minute prices are computed from public aggregates, while individual positions and financial balances remain behind cryptographic commitments. Zero-knowledge proofs, signed receipts and member recourse link private transactions to the public settlement conditions and prevent unverified or partial on-chain settlement. Controlled execution shows feasible proving time and daily gas requirements for the ten-household case, as well as the expected loss of liveness under insufficient QBFT validator quorum.

The fixed-profile exercise demonstrates the gains from local matching under an assumed tariff spread: 34.4\% of demand is served locally and the grid-only net bill falls by 30.5\% in the four-day case study. A separate strategic scheduling exercise finds household equilibria recovering approximately 83--88\% of the additional planner gain, while highlighting the multiplicity of equilibria. The protocol therefore offers a reproducible basis for studying how market design, technical verification and participant incentives interact. A larger-scale, geographically distributed deployment and a completed jurisdiction-specific legal analysis are the natural next step for making this prototype operational.

\end{document}